\documentclass[aps,pre,floats,twocolumn,superscriptaddress,floatfix]{revtex4}
\usepackage{amsmath,amsfonts,amssymb}
\usepackage[pdftex]{graphicx}

\begin{document}

\title{Navigability of complex networks}

\author{Mari{\'a}n Bogu{\~n}{\'a}}

\affiliation{Departament de F{\'\i}sica Fonamental, Universitat de
  Barcelona, Mart\'{\i} i Franqu\`es 1, 08028 Barcelona, Spain}

\author{Dmitri Krioukov}

\author{kc claffy}

\affiliation{Cooperative Association for Internet Data
Analysis (CAIDA), University of California, San Diego (UCSD), 9500
Gilman Drive, La Jolla, CA 92093, USA}

\begin{abstract}
Routing information through networks is a universal phenomenon
in both natural and manmade complex systems.  When each node
has full knowledge of the global network connectivity, finding short
communication paths is merely a matter of distributed computation.
However, in many real networks nodes communicate efficiently even
without such global intelligence.
Here we show that the peculiar structural characteristics of many complex networks support efficient communication without global knowledge.
We also describe a general mechanism that explains this connection between
network structure and function. This mechanism relies on the
presence of a metric space hidden behind an observable network.
Our findings suggest that real networks in nature have
underlying metric spaces that remain undiscovered. Their discovery
would have practical applications ranging from routing in the Internet
and searching social networks, to studying information flows in neural,
gene regulatory networks, or signaling pathways.
\end{abstract}

\maketitle

\section{Introduction}

Networks are ubiquitous in all domains of science and technology,
and permeate many aspects of daily human
life~\cite{Albert:2002,newman03c-review,Dorogovtsev:2003,Boccaletti:2006},
especially upon the rise of the information technology
society~\cite{Castells:2006,RomusVespasbook}. Our growing
dependence on them has inspired a burst of activity in the
new field of network science, keeping researchers motivated to
solve the difficult challenges that networks offer.
Among these, the relation between network structure and function is
perhaps the most important and fundamental. Transport is one of
the most common functions of networked systems. Examples can be
found in many domains: transport of energy in metabolic networks, of
mass in food webs, of people in transportation systems, of
information in cell signalling processes, or of bytes across
the Internet.

In many of these examples, routing --or signalling of information
propagation paths through a complex network maze-- plays a
determinant role in the transport properties of the system,
in particular in such systems as the Internet or airport networks
that have transport as their primary function. The
observed efficiency of this routing process in real networks poses an
intriguing question: how is this efficiency achieved?
When each element of the system has a
full view of the global network topology, finding short routes
to target destinations is a well-understood computational process.
However, in many networks observed in nature, including those in
society and biology (signalling pathways, neural networks, etc.),
nodes efficiently find intended communication targets even though
they do not possess any global view of the system. For example,
neural networks would not function so well if they could not route
specific signals to appropriate organs or muscles in the body,
although no neurone has a full view of global inter-neurone
connectivity in the brain.

In this work, we identify a general mechanism that explains routing
conductivity, or navigability of real networks based on the concept
of similarity between nodes~\cite{WatDoNew02,GiNe02,menczer02-pnas,LeHoNe06,CraCo08,ClMo08}.
Specifically, intrinsic characteristics of nodes define a measure of
similarity between them, which we abstract as a hidden distance.
Taken together, hidden distances define a {\it hidden metric space}
for a given network. Our recent work shows that these spaces explain
the observed structural peculiarities of several real networks, in
particular social and technological ones~\cite{SerKriBog07}. Here
we show that this underlying metric structure can be used to guide
the routing process, leading to efficient communication without
global information in arbitrarily large networks. Our analysis
reveals that, remarkably, real networks satisfy the topological
conditions that maximise their navigability within this framework.
Therefore, hidden metric spaces offer explanations of two open
problems in complex networks science: the communication efficiency
networks so often exhibit, and their unique structural
characteristics.

\section{Node similarity and hidden metric spaces}

Our work is inspired by the seminal work of sociologist Stanley
Milgram on the small world problem. The small world paradigm refers
to the existence of short chains of acquaintances among individuals
in societies~\cite{TraMi69}. At Milgram's time, direct proof of
such a paradigm was impossible due to the lack of large databases of
social contacts, so Milgram conceived an experiment to analyse the
small world phenomenon in human social networks.
Randomly chosen individuals in the United States were asked to
route a letter to an unknown recipient using only friends or
acquaintances that, according to their judgement, seemed most likely
to know the intended recipient.  The outcome of the experiment revealed
that, without any global network knowledge, letters reached the target
recipient using, on average, $5.2$ intermediate people, demonstrating that social acquaintance networks were indeed small worlds.

\begin{figure*}
\begin{center}
\includegraphics[width=6in]{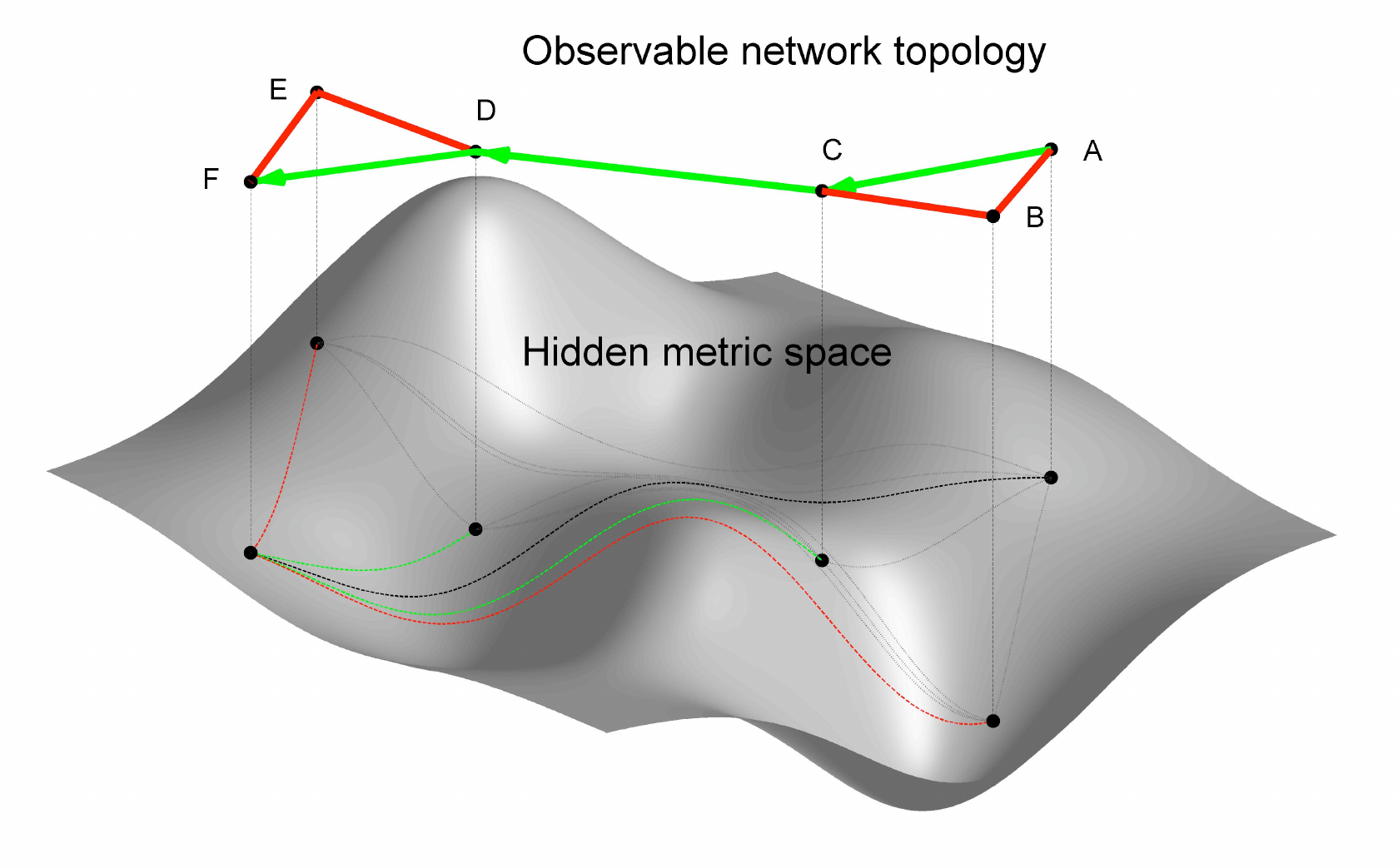}
\end{center}
\vspace{-0.5cm} \caption{{\bf How hidden metric spaces influence the
structure and function of complex networks.} The smaller the
distance between two nodes in the hidden metric space, the more
likely they are connected in the observable network topology. If
node $A$ is close to node $B$, and $B$ is close to $C$,
then $A$ and $C$ are necessarily close because of the
triangle inequality in the metric space. Therefore, triangle $ABC$
exists in the network topology with high probability, which explains
the strong clustering observed in real complex networks. The hidden
space also guides the greedy routing process: if node $A$ wants to
reach node $F$,
it checks the hidden distances between $F$ and its two neighbours $B$
and $C$. Distance $CF$ (green dashed line) is smaller than $BF$ (red
dashed line), therefore $A$ forwards information to $C$. Node $C$
then performs similar calculations and selects its neighbour $D$ as
the next hop on the path to $F$. Node $D$ is directly connected to
$F$. The result is path $A \to C \to D \to F$ shown by green edges
in the observable topology. \label{fig:hms}}
\end{figure*}

The small world property can be easily induced by adding a small
number of random connections to a ``large world''
network~\cite{Watts:1998}. More striking is the fact that social
networks are navigable without global information. Indeed, the only
information that people used to make their routing decisions in
Milgram's experiment was a set of descriptive attributes of the
destined recipient, such as place of living and occupation. People
then determined who among their contacts was ``socially closest'' to
the target.  The success of the experiment indicates that social
distances among individuals --even though they may be difficult to
define mathematically-- play a role in shaping the network
architecture and that, at the same time, these distances can be used
to navigate the network.  However, it is not clear how this coupling
between the structure and function of the network leads to efficiency
of the search process, or what the minimum structural requirements are
to facilitate such efficiency~\cite{kleinberg00-nature}.

In this work, we show how network navigability depends on the
structural parameters characterising the two most prominent and
common properties of real complex networks: (1)~scale-free (power-law) node
degree distributions characterising the heterogeneity in the number
of connections that different nodes have, and (2)~clustering, a
measure of the number of triangles in the network topology. We
assume the existence of a hidden metric space, an underlying
geometric frame that contains all nodes of the network, shapes its
topology, and guides routing decisions, as illustrated in
Fig.~\ref{fig:hms}. Nodes are connected in the observable topology,
but a full view of their global connectivity is not available at any
node. Nodes are also positioned in the hidden metric space and
identified by their co-ordinates in it. Distances between nodes
in this space abstract their similarity~\cite{WatDoNew02,GiNe02,menczer02-pnas,LeHoNe06,CraCo08,ClMo08}.
These distances
influence both the observable topology and routing function:
(1)~the smaller the distance between two nodes in the
hidden space, i.e., the more similar the two nodes, the more likely
they are connected in the observable topology; (2)~nodes also use hidden
distances to select, as the next hop, the neighbour closest to the
destination in the hidden space. Kleinberg introduced the term
{\it greedy routing} to describe this forwarding
process~\cite{kleinberg00-nature}.
Greedy routing and its modifications have been studied extensively
in recent computer science
literature~\cite{kleinberg00-stoc,kleinberg01,MaNaWe04,MaNg04,NgMa05,NgMa08,SiJe05,LeSha04,fraigniaud05,FraGaPa06,FraGa07,FraGa08,ChaFra08}
(see also Kleinberg's review~\cite{kleinberg06-review-pnas}
and references therein). However,
most of these works do not
study greedy routing on scale-free topologies, which are known
as the common signature of many large-scale self-evolving
complex networks~\cite{Albert:2002,newman03c-review,Dorogovtsev:2003}.

We use the class of network models developed in recent
work~\cite{SerKriBog07}. They generate networks with topologies
similar to those of real networks --small-world, scale-free, and
with strong clustering-- and, simultaneously, with hidden metric
spaces lying underneath. The simplest model in this class (the details are in Appendix~\ref{sec:app:model}) uses a
one-dimensional circle as the underlying metric space, in which nodes
are uniformly distributed. The model first assigns to each node its
expected degree $k$, drawn from a power-law degree distribution
$P(k) \sim k^{-\gamma}$, with $\gamma>2$, and then connects each
pair of nodes with connection probability $r(d;k,k')$ that depends
both on the distance $d$ between the two nodes in the circle and
their assigned degrees $k$ and $k'$,
\begin{eqnarray}\label{eq:all}
r(d;k,k') &\equiv& r(d/d_c) = \left(1+d/d_c\right)^{-\alpha},\\
&&\text{where $\alpha>1$ and $d_c \sim kk'$\nonumber,}
\end{eqnarray}
which means that the probability of link connection between two
nodes in the network decreases with the hidden distance between them
(as $\sim d^{-\alpha}$) and increases with their degrees (as $\sim
(kk')^\alpha$).

These two properties have a clear interpretation. The connection
cost increases with hidden distance, thus discouraging long-range
links. However, in making connections, rich (well-connected,
high-degree) nodes care less about distances (connection costs) than poor
nodes. Further, the {\em characteristic distance scale} $d_c$
provides a coupling between node degrees and hidden distances, and
ensures the following three topological characteristics that we
commonly see in real networks. First, pairs of richly connected,
high-degree nodes --{\it hubs}-- are connected with high probability
regardless of the hidden distance between them because their
characteristic distance $d_c$ is so large that any actual distance
$d$ between them will be short in comparison: regardless of $d$,
connection probability $r$ in Eq.~(\ref{eq:all}) is close to 1 if
$d_c$ is large. Second, pairs of low-degree nodes will not be
connected unless the hidden distance $d$ between them is short
enough to compare with the small value of their characteristic
distance $d_c$. Third, following similar arguments, pairs composed
of hubs and low-degree nodes are connected only if they are located
at moderate hidden distances.

The parameter $\alpha$ in Eq.~(\ref{eq:all}) determines the
importance of hidden distances for node connections. The larger
$\alpha$, the more preferred are connections between nodes close in
the hidden space. Consequently, the triangle inequality in the
metric space leads to stronger clustering in the network,
cf.~Fig.~\ref{fig:hms}. Clustering has a clear interpretation in our
approach as a reflection of the network's metric strength: the more
powerful is the influence of the network's underlying metric space
on the observable topology, the more strongly it is clustered.

Although our toy model is not designed to exactly
match any specific real network, it generates graphs that are
surprisingly similar to some real networks, such as the Internet at
the autonomous system level or the USA airport network. See
Appendix~\ref{sec:app:model-vs-reality} for details.

\section{Navigability of modelled networks}

\begin{figure}
\begin{center}
\includegraphics[width=3.5in]{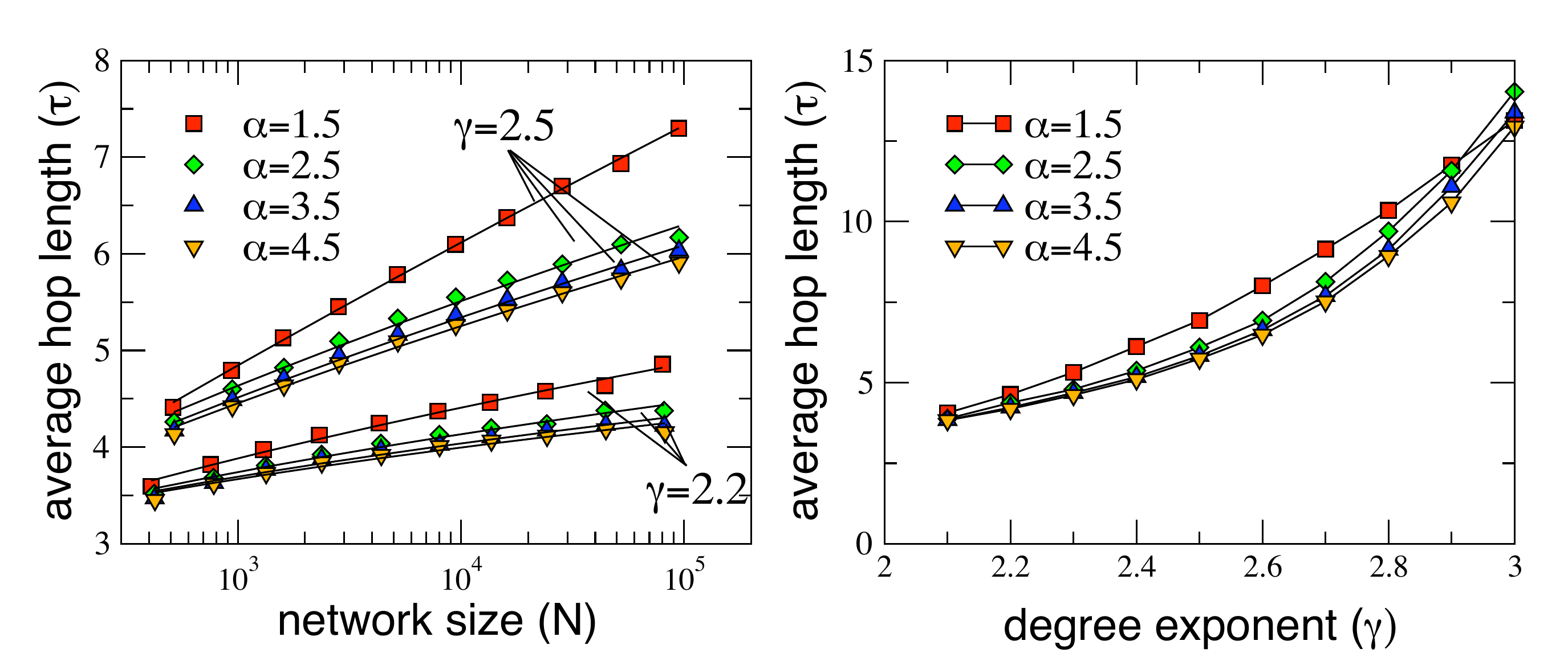}
\end{center}
\vspace{-0.5cm} \caption{{\bf Average length of
greedy-routing paths.} The left plot shows the average hop length of
successful paths, $\tau$, as a function of the network size $N$ for
different values of $\gamma$ and $\alpha$. Results for values of
$\gamma>2.5$ look similar but with longer paths and are
 omitted for clarity. In all cases, the path length
grows polylogarithmically with the network size: the observed values
of $\tau$ are fit well by $\tau(N)=A[\log{N}]^{\nu}$ (solid lines),
where $A$ and $\nu$ are some constants. The right plot shows $\tau$
as a function of $\gamma$ and $\alpha$ for networks of fixed size
$N\approx10^5$.  The effect of the two parameters on average path
length is straightforward: paths are shorter for smaller exponents
$\gamma$ and stronger clustering (larger $\alpha$'s).  \label{greedy2}}
\end{figure}

\begin{figure}
\begin{center}
\includegraphics[width=3.5in]{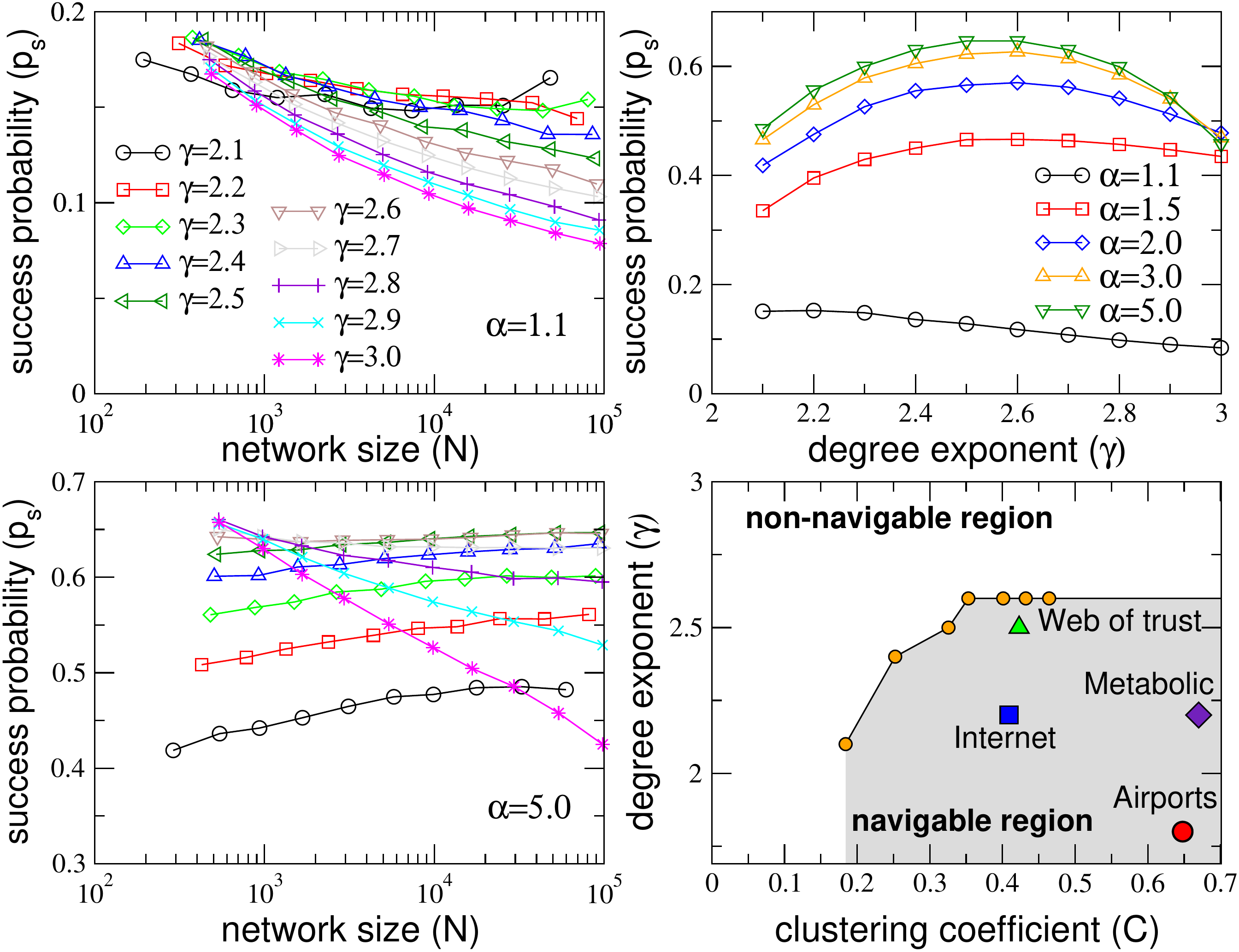}
\end{center}
\vspace{-0.5cm} \caption{{\bf Success probability of greedy
routing.} Left plots:
success probability $p_s$
as a function of network size $N$ for different values of $\gamma$ with
weak (top) and strong (bottom) clustering. The top-right plot shows
$p_s$ as a function of $\gamma$ and $\alpha$ for networks of fixed
size $N \approx 10^5$. In the bottom-right plot, parameter $\alpha$
is mapped to clustering coefficient $C$~\cite{Watts:1998} by
computing $C$ for each network with given $\gamma$ and $\alpha$. For
each value of $C$, there is a critical value of $\gamma=\gamma_c(C)$
such that the success ratio in networks with this $C$ and
$\gamma>\gamma_c(C)$ decreases with the network size ($p_s(N)
\xrightarrow[N\to\infty]{} 0$), while $p_s(N)$ reaches a constant
value for large $N$ in networks with $\gamma<\gamma_c(C)$. The solid
line in the plot shows these critical values $\gamma_c(C)$,
separating the low-$\gamma$, high-$C$ navigable region, in which
greedy routing remains efficient in the large-graph limit, from the
high-$\gamma$, low-$C$ non-navigable region, where the efficiency of
greedy routing degrades for large networks.  The plot
labels measured values of $\gamma$ and $C$ for several real
complex networks. {\it Internet} is the global Internet topology
of autonomous systems as seen by the Border
Gateway Protocol (BGP)~\cite{MaKrFo06}; {\it Web of trust} is the
Pretty Good Privacy (PGP) social network of mutual trust
relationships~\cite{BoPa04a}; {\it Metabolic} is the network of
metabolic reactions of {\it E.~coli}~\cite{JeToAlOlBa00}; and {\it
Airports} is the network of the public air transportation
system~\cite{Barrat:2004b}. \label{p_s}}
\end{figure}

We use the model to generate scale-free networks with different
values of power-law degree distribution exponent $\gamma$ and
clustering strength $\alpha$, covering the observed values in a vast
majority of documented complex
networks~\cite{Albert:2002,newman03c-review,Dorogovtsev:2003}. We
then simulate greedy routing for a large sample of paths on all
generated networks, and compare the following two navigability
parameters: 1)~the average hop length $\tau$ from source to
destination of successful greedy-routing paths, and 2)~the success
ratio $p_s$, defined as the percentage of successful paths.
Unsuccessful paths are paths that get stuck at nodes without
neighbours closer to the destination in the hidden space than
themselves. These nodes usually have small degrees. See Appendix~\ref{sec:app:simulations}
for simulation details.

Fig.~\ref{greedy2}~shows the impact of the network's degree
distribution and clustering on the average length $\tau$ of greedy
routing paths. We observe a straightforward dependency: paths
are shorter for smaller exponents $\gamma$ and stronger clustering
(larger $\alpha$'s). The dependency of the success ratio
(the fraction of successful paths) $p_s$ on
the two topology parameters $\gamma$ and $\alpha$ is more
intertwined. Fig.~\ref{p_s} shows that the effect of one parameter,
$\gamma$, on the success ratio depends on the other parameter,
the level of clustering.
If clustering is weak (low $\alpha$), the percentage of successful
paths decays with network size~$N$ regardless of the value of $\gamma$
(Fig.~\ref{p_s} top-left). However, with strong clustering (large
$\alpha$), the percentage of successful paths increases with $N$ and
attains a maximum for large networks if $\gamma \lesssim 2.6$,
whereas it degrades for large networks if $\gamma>2.6$
(Fig.~\ref{p_s} bottom-left). Fig.~\ref{p_s} top-right shows this
effect for networks of the same size ($N=10^5$) with different
$\gamma$ and $\alpha$. The value of $\gamma=2.6 \pm 0.1$ maximises
the number of successful paths once clustering is above a threshold,
$\alpha \geq 1.5$. These observations mean that for a fixed
clustering strength, there is a critical value of the exponent
$\gamma$ (Fig.~\ref{p_s} bottom-right) below which networks remain
navigable as their size increases, but above which their
navigability deteriorates with their size.

In summary, strong clustering improves both navigability metrics. We
also find a delicate trade-off between values of $\gamma$ close to
$2$ minimising path lengths, and higher values -- not exceeding
$\gamma \approx 2.6$ -- maximising the percentage of successful
paths. We explain these findings in the next section, but we note
here that qualitatively, {\em this navigable parameter region
contains a majority of complex networks observed in
reality}~\cite{Albert:2002,newman03c-review,Dorogovtsev:2003}, as
confirmed in Fig.~\ref{p_s} (bottom-right), where we juxtapose few
paradigmatic examples of communication, social, biological, and
transportation networks vs.\ the identified navigable region of
clustering and degree distribution exponent.
Interestingly, power grids, which propagate electricity
rather than route information, are neither scale-free nor clustered~\cite{Watts:1998,Amaral:2000}.

\section{Air travel by greedy routing as an explanation}

\begin{figure*}
\begin{center}
\includegraphics[width=6.5in]{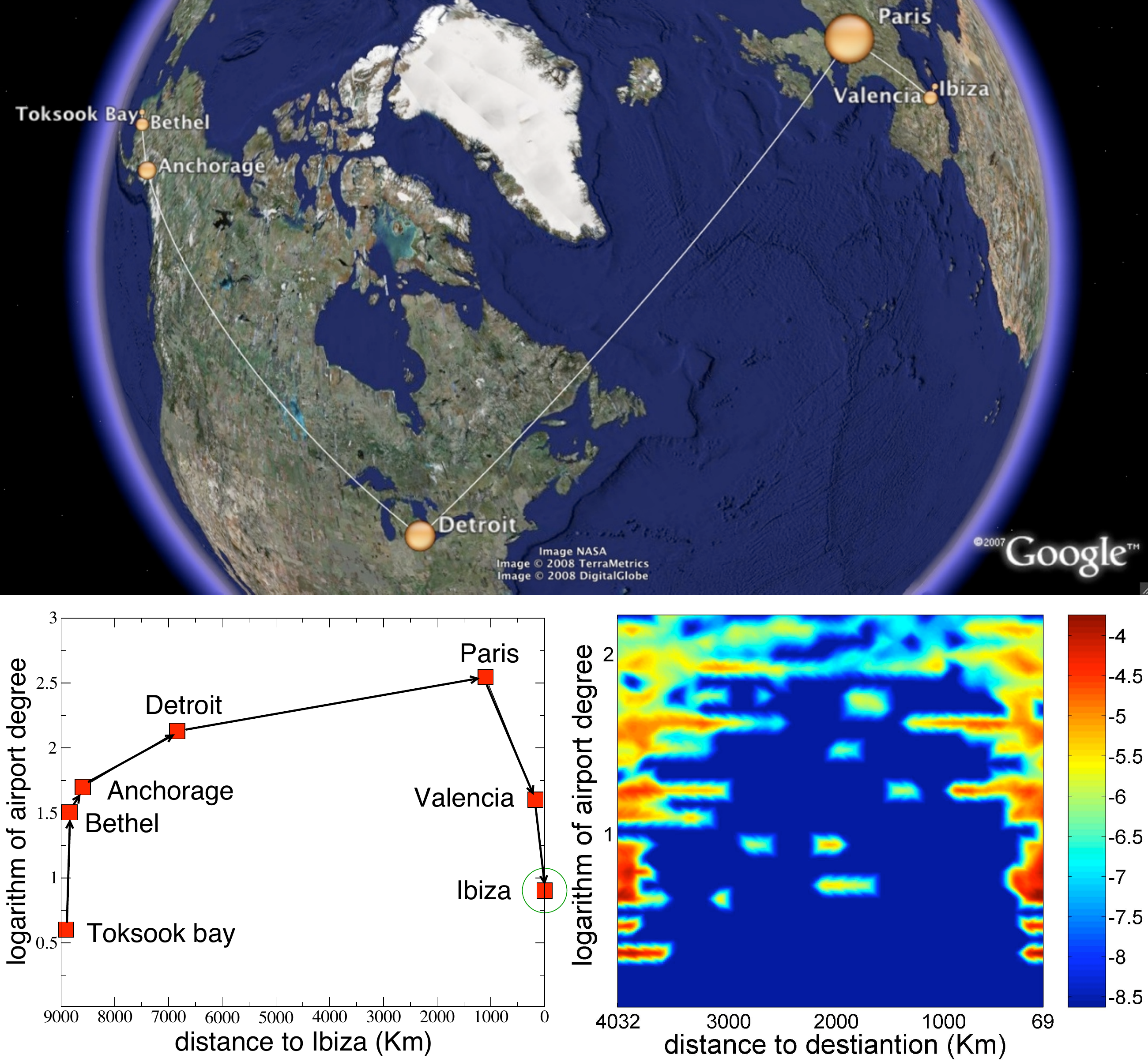}
\end{center}
\vspace{-0.5cm} \caption{{\bf Greedy routing in the
airport network. Top: the structure of the single greedy-routing
path from Toksook Bay to Ibiza.} At each intermediate airport, the
next hop is the airport closest to Ibiza geographically. Sizes of
symbols representing the airports are proportional to the logarithm
of their degrees. {\bf The bottom left} figure shows the changing
distance to Ibiza (in the $x$ axis) vs.\ the degree of the visited
airports ($y$ axis, in logarithmic scale). {\bf
Bottom right: the structure of greedy-routing paths between a
collection of airports in the USA~\cite{USAN}.} We include an
airport pair in the collection if the distance between the airports
is between $3900$ and $4100$ kilometers. The number of airport pairs
in this collection is $7620$. We use colour to indicate how often
paths in the collection go through an airport of a given degree
located at a given geographical distance from the destination:
blue/red indicates exponentially less/more visits to those airports,
or more specifically, the color is the logarithm of the normalised
density of visited airports.
\label{airtravel}}
\end{figure*}

We illustrate the greedy routing function, and the structure of
networks conductive to such routing, with an example of passenger
air travel. Suppose we want to travel from Toksook Bay, Alaska, to
Ibiza, Spain, by the public air transportation network. Nodes in
this network are airports, and two airports are connected if there
is at least one flight between them. We travel according to the
greedy routing strategy using geography as the underlying metric
space. At each airport we choose the next-hop airport geographically
closest to the destination. Under these settings, our journey goes
first to Bethel, then to Anchorage, to Detroit, over the Atlantic to
Paris, then to Valencia and finally to Ibiza, see
Fig.~\ref{airtravel}. The sequence and sizes of airport hops reveal
the structure of our greedy-routing path. The path proceeds from a
small airport to a local hub at a small distance, from there to a
larger hub at a larger distance, and so on until we reach Paris. At
that point, when the distance to the destination becomes
sufficiently small, greedy routing leads us closer to our final
destination by choosing not another hub, but a less connected
neighbouring airport.

We observe that the navigation process has two, somewhat symmetric
phases. The first phase is a coarse-grained search, travelling longer
and longer distances per hop toward hubs, thus ``zooming out'' from
the starting point.
The second phase corresponds to a fine-grained search, ``zooming
in'' onto the destination. The turning point between the two phases
appears naturally: once we are in a hub near the destination, the
probability that it is connected to a bigger hub closer to the
destination sharply decreases, but at this point we do not need hubs
anyway, and greedy routing directs us to smaller airports at shorter
distances next to the destination.

This {\it zoom out/zoom in} mechanism works efficiently only if the
coupling between the airport network topology and the underlying
geography
satisfies the following two conditions: the {\em sufficient hubs}
condition and the {\em sufficient clustering} condition. The first
condition ensures that a network has enough hub airports
(high-degree nodes) to provide an increasing sequence during the
zoom out phase. This condition is fulfilled by the real airport
network and by other scale-free networks with small values of degree
distribution exponent $\gamma$, because the smaller the $\gamma$,
the larger the proportion of hubs in the network.

However, the presence of many hubs does not ensure that greedy
routing will use them. Unlike humans, who can use their knowledge
of airport size to selectively travel via hub airports, greedy
routing uses only one constraint at each hop: minimise distance
to the destination.
Therefore, the network topology must satisfy the second condition,
which ensures that Bethel is larger than Toksook Bay, Anchorage
larger than Bethel, and so on. More generally, this condition is
that the next greedy hop from a remote low-degree node likely has a
higher degree, so that greedy paths typically head first toward the
highly connected network core.
But the network metric strength is exactly the required property:
preference for connections between nodes nearby in the
hidden space means that low-degree nodes are less likely
to have connectivity to distant low-degree nodes; only high-degree
nodes can have long-range connection that greedy routing will effectively select.
The stronger this coupling between the metric space and topology
(the higher $\alpha$ in Eq.~(\ref{eq:all})), the stronger the
clustering in the network.

To illustrate, imagine an airport network without
sufficient clustering, one where the airport closest to our
destination (Ibiza) among all airports connected to
our current node (Toksook Bay, Alaska) is not Bethel,
which is bigger than Toksook Bay, but
Nightmute, Alaska, a nearby airport of comparable size
to Toksook Bay. As greedy routing first
leads us to Nightmute, then to another small nearby airport,
and then to another, we can no longer get to Ibiza
in few hops. Worse, travelling via these numerous small
airports, we could reach one
with no connecting flights heading closer to Ibiza. Our
greedy routing would be stuck at this airport with an unsuccessful path.

These factors explain why the most navigable topologies
correspond to scale-free networks with small exponents of the degree
distribution, i.e., a large number of hubs, and with strong
clustering, i.e., strong coupling between the hidden geometry and
the observed topology.

\section{The structure of greedy-routing paths}

\begin{figure}
\begin{center}
\includegraphics[width=3.5in]{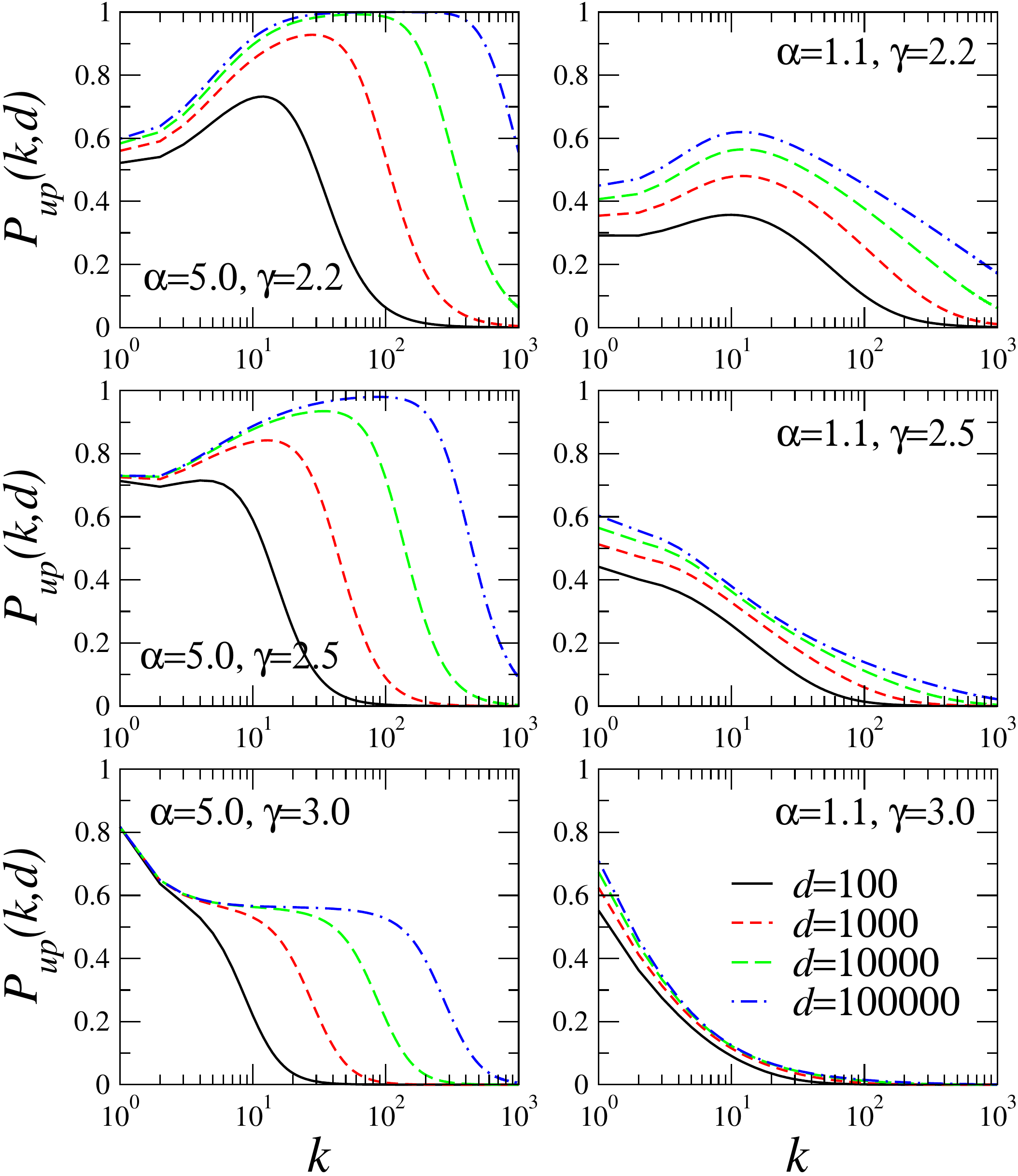}
\end{center}
\vspace{-0.5cm} \caption{{\bf Probability that greedy routing
travels to higher-degree nodes.} More precisely, the probability
$P_{up}(k,d)$ that the greedy-routing next hop after a node of
degree $k$ located at distance $d$ from a destination has higher
degree $k' \geq k$ and is closer to the destination. The distance
legend in the right-bottom plot applies to all the plots. The
results are for the large-graph limit $N\to\infty$.
\label{fig:P_up}}
\end{figure}

\begin{figure}
\begin{center}
\includegraphics[width=3.5in]{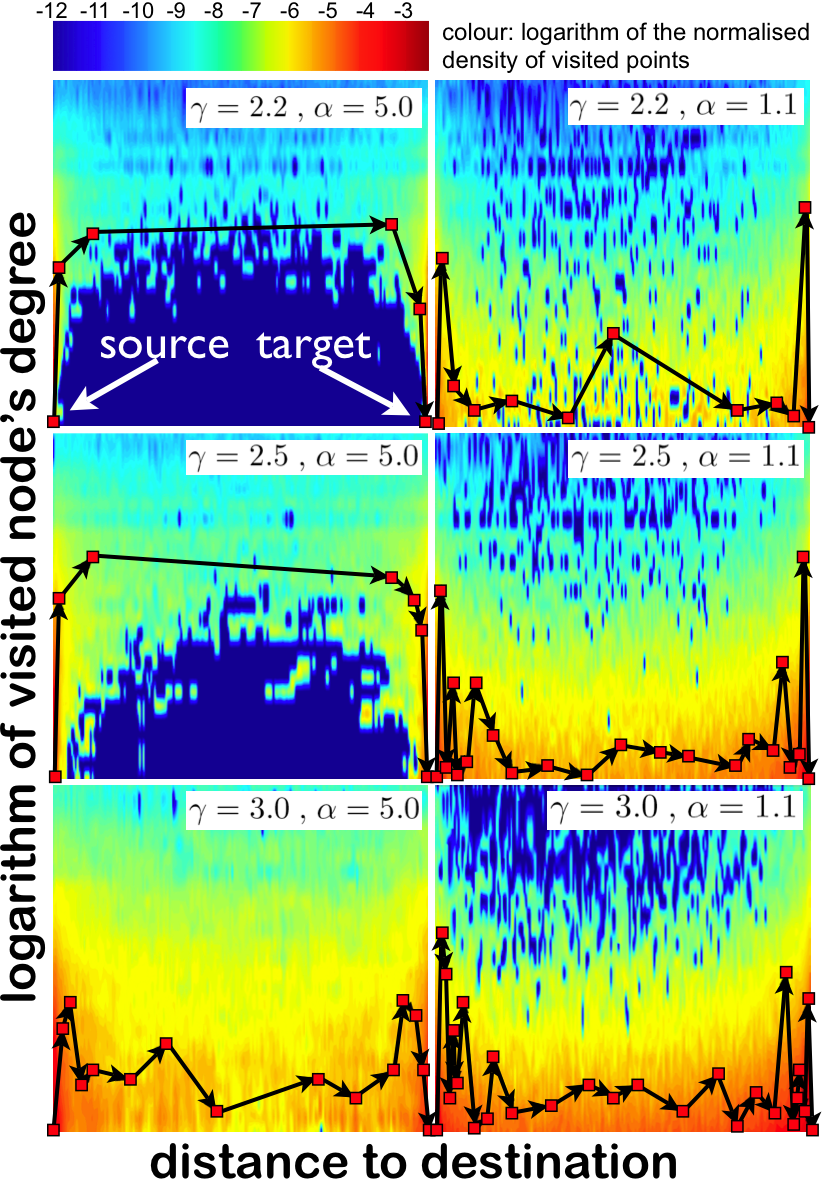}
\end{center}
\vspace{-0.5cm} \caption{ {\bf The structure
of greedy-routing paths.}
We visualise the results of our simulation of greedy routing
in modelled networks with different
values of $\gamma$ and $\alpha$ observed in real complex networks.
The hidden distance between the starting point and the destination
is always approximately $10^4$, and the network size $N$ and
number of attempted paths is always $10^5$ for
each $(\gamma,\alpha)$ combination, but the number of successful
paths and path hop-lengths vary, cf.~Figs.~\ref{greedy2},\ref{p_s}.
All paths start and end at low-degree nodes located, respectively,
in the left- and right-bottom corners of the diagrams (see top left
plot). For each $(\gamma,\alpha)$ we depict a
single typical path in black and, as in Fig.~\ref{airtravel}, use colour to indicate how often
paths included a node of a given degree located at a given distance from
the destination. The simulations confirm that only when $\gamma$ is small and
$\alpha$ is large does the average path structure follow the
zoom-out/zoom-in pattern that characterises successful greedy
routing in real networks, e.g., in the airport network in Fig.~\ref{airtravel}.
\label{fig:path_hierarchy}}
\end{figure}

We observe the discussed zoom-out/zoom-in mechanism in analytical
calculations and numerical simulations. Specifically, we calculate
(in Appendix~\ref{sec:app:propagator}) the probability that the next hop from a
node of degree $k$ located at hidden distance $d$ from the
destination has a larger degree $k'>k$, in which case the path moves
toward the high-degree network core, see Fig.~\ref{fig:P_up}. In the
most navigable case, with small degree-distribution exponent and
strong clustering, the probability of increasing the node degree
along the path is high at low-degree nodes, and sharply decreases to
zero after reaching a node of a critical degree value, which
increases with distance~$d$. This observation implies that
greedy-routing paths first propagate up to higher-degree nodes in
the network core and then exit the core toward low-degree
destinations in the periphery. In contrast, with low clustering,
paths are less likely to find higher-degree nodes regardless of the
distance to the destination. This path structure violates the
zoom-out/zoom-in pattern required for efficient navigation.

Fig.~\ref{fig:path_hierarchy} shows the structure of greedy-routing
paths in simulations, further confirming our analysis. We again see
that for small degree-distribution exponents and strong clustering
(upper left and middle left), the routing process quickly finds a
way to the high-degree core, makes a few hops there, and then
descends to a low-degree destination. In the other, non-navigable
cases, the process can almost never get to the core of high-degree
nodes. Instead, it wanders in the low-degree periphery increasing
the probability of getting lost at low-degree nodes.

\section{Discussion}

Our main motivation for this work comes from
long-standing scalability problems with the Internet routing
architecture~\cite{iab-raws-report-phys}. To route information
packets to a given destination, Internet routers must communicate to
maintain a coherent view of the global Internet topology. The
constantly increasing size and dynamics of the Internet thus leads
to immense and quickly growing communication and information
processing overhead, a major bottleneck in routing
scalability~\cite{KrKc07} causing concerns among Internet experts
that the existing Internet routing architecture may not sustain even
another decade~\cite{iab-raws-report-phys}. Discovery of the
Internet's hidden metric space would remove this bottleneck,
eliminating the need for the inherently unscalable communication of
topology changes.
Instead routers would be able to just forward packets
greedily to the destination based on hidden distances.

In a similar manner, reconstruction of hidden metric spaces underlying
other real networks may prove practically useful. For example,
in social or communication networks (e.g., the Web, overlay, or
online social networks) hidden spaces would yield efficient strategies
for searching specific individuals or content based only
on local knowledge. The metric spaces
hidden under some biological networks (such as neural, gene regulatory networks,
signalling or even protein folding~\cite{RaGna07} pathways) can become a powerful tool in studying the
structure of information or signal flows in these networks,
enabling investigation of such processes without detailed
global knowledge of the network structure or organisation.

The natural question we thus face is how to proceed
toward discovery of the explicit structure of hidden metric spaces
underlying real networks. We do not expect spaces underlying different
networks to be exactly the same. For example, the similarity spaces
of Web pages~\cite{menczer02-pnas} and Wikipedia editors~\cite{CraCo08} likely
differ. However, the main contribution of this work
establishes the {\em general\/} mechanisms behind navigability of
scale-free, strongly clustered topologies that characterise many
different real networks. The next step is to find the common properties of hidden
spaces that render them congruent with these mechanisms.
Specifically, we are interested in what geometries of hidden spaces
lead to such congruency~\cite{KrPa08}.

In general, we believe that the present and future work on hidden
metric spaces and network navigability will deepen our understanding
of the fundamental laws describing relationships between structure
and function of complex networks.

\begin{acknowledgments}

We thank M. \'Angeles Serrano for useful comments and discussions. This
work was supported in part by DGES grant FIS2007-66485-C02-02,
Generalitat de Catalunya grant No. SGR00889, the Ram\'on y Cajal
program of the Spanish Ministry of Science, by NSF CNS-0434996
and CNS-0722070, by DHS N66001-08-C-2029, and by Cisco Systems.

\end{acknowledgments}

\appendix

\section{A model with the circle as a hidden metric space.}
\label{sec:app:model}

In our model we place all nodes on a circle by assigning them a
random variable $\theta$, i.e., their polar angle, distributed
uniformly in $[0,2\pi)$. The circle radius $R$ grows linearly with
the total number of nodes $N$, $2 \pi R= N$, in order to keep the
average density of nodes on the circle fixed to $1$. We next assign
to each node its expected degree $\kappa$ drawn from some
distribution $\rho(\kappa)$. The connection probability between two
nodes with hidden co-ordinates $(\theta,\kappa)$ and
$(\theta',\kappa')$ takes the form
\begin{equation}\label{eq:r-factoring}
r(\theta,\kappa;\theta',\kappa')= \left(
1+\frac{d(\theta,\theta')}{\mu \kappa \kappa'}
\right)^{-\alpha},\quad \mu=\frac{(\alpha-1)}{ 2
\langle k \rangle},
\end{equation}
where $d(\theta,\theta')$ is the geodesic distance between the two
node on the circle, while $\langle k \rangle$ is the average degree.
One can show that the average degree of nodes with hidden variable
$\kappa$, $\bar{k}(\kappa)$, is proportional to
$\kappa$.\cite{BoPa03}~~ This proportionality guarantees that
the shape of the node degree distribution $P(k)$ in generated
networks is approximately the same as the shape of $\rho(\kappa)$.
The choice of $\rho(\kappa) = (\gamma-1)\kappa_0^{\gamma-1}
\kappa^{-\gamma}$, $\kappa>\kappa_0\equiv (\gamma-2) \langle k
\rangle/(\gamma-1)$, $\gamma > 2$,  generates random networks with a
power-law degree distribution of the form $P(k) \sim k^{-\gamma}$,
where
$\gamma$ is a model parameter that regulates
the heterogeneity of the degree distribution in the network. This parameter
abstracts the heterogeneity of node degrees in real networks,
where degree distributions may not perfectly follow power laws,
or may exhibit various forms of high-degree cut-offs~\cite{Giot03,MaKrFo06}.
The specific effects are less important than the overall
measure of heterogenity.
We note that instead of a circle in our model we could use
any isotropic space of any dimension~\cite{SerKriBog07}.

\section{Numerical simulations.}
\label{sec:app:simulations}

Our model has three independent parameters: exponent $\gamma$ of
power-law degree distributions, clustering strength $\alpha$, and
average degree $\langle k \rangle$. We fix the latter to $6$, which
is roughly equal to the average degree of some real networks of
interest~\cite{MaKrFo06,BoPa04a}, and vary $\gamma\in[2.1,3]$ and
$\alpha\in[1.1,5]$, covering their observed ranges in documented
complex
networks~\cite{Albert:2002,newman03c-review,Dorogovtsev:2003}. For
each $(\gamma,\alpha)$ pair, we produce networks of different sizes
$N\in[10^3,10^5]$ generating, for each $(\gamma,\alpha,N)$, a number
of different network instances---from $40$ for large $N$ to $4000$
for small $N$. In each network instance $G$, we randomly select
$10^6$ source-destination pairs $(a,b)$ and execute the
greedy-routing process for them starting at $a$ and selecting, at
each hop $h$, the next hop as the $h$'s neighbour in $G$ closest to
$b$ in the circle. If for a given $(a,b)$, this process visits the
same node twice, then the corresponding path leads to a loop and is
unsuccessful. We then average the measured values of path hop
lengths $\tau$ and percentage of successful paths $p_s$ across all
pairs $(a,b)$ and networks $G$ for the same $(\gamma,\alpha,N)$.
Note that we are not concerned with the absolute
values of the success ratio
$p_s$. Instead we use it as {\em a measure\/} of navigability
to compare networks with different $(\gamma,\alpha,N)$. For this
purpose we could use the success ratio of any (improved)
modification of standard greedy routing.

\section{Shortest path {\it vs.} shortest time.}
\label{sec:app:shortest}

All results derived in the present paper are about finding
short paths across a network topology. The total physical time
from source to destination is implicitly assumed to be proportional
to the number of hops. In real transportation systems, e.g. the
Internet or the airport network, the finite capacity of nodes
implies that the end-to-end path latency may be longer
when intermediate nodes are congested.  While our results
most cleanly apply to uncongested systems, there are
obvious modifications, such as choosing the second or
third nearest rather than the nearest neighbor, that
could still find nearly shortest paths while reducing
and balancing load on the system.

\section{The model {\it vs.} real networks: the Autonomous System level map of the Internet and the US Airport Network}
\label{sec:app:model-vs-reality}

The model we use in this work is not meant to reproduce any
particular system but to generate a set of general properties, like
heterogeneous degree distributions, high clustering, and a metric
structure lying underneath. Yet, despite its simplistic assumptions,
the model generates graphs that are surprisingly close to some real
networks of interest, in particular the Internet at the Autonomous
System level (AS)~\cite{MaKrFo06,dimes-ccr} and the network of
airline connections among airports within the United States during
2006 (USAN)~\cite{USAN}. In the case of the Internet, we use two
different data sets, the Internet as viewed by the Border Gateway
Protocol (BGP)~\cite{MaKrFo06} and the DIMES
project~\cite{dimes-ccr}. The BGP (DIMES) network has a size of
$N\sim 17446$ ($N=19499$) ASs, average degree $\langle k
\rangle=4.7$ ($\langle k \rangle=5$), and average clustering
$C=0.41$ ($C=0.6$). The US Airport Network is composed of US
airports connected by regular flights (with more than 1000
passengers per year) during the year $2006$. This results in a
network of $N=599$ airports, average degree $\langle k \rangle \sim
10.8$ and average clustering coefficient $C=0.72$.

\begin{figure}[t]
\begin{center}
\begin{tabular}{c}
\includegraphics[width=8.5cm]{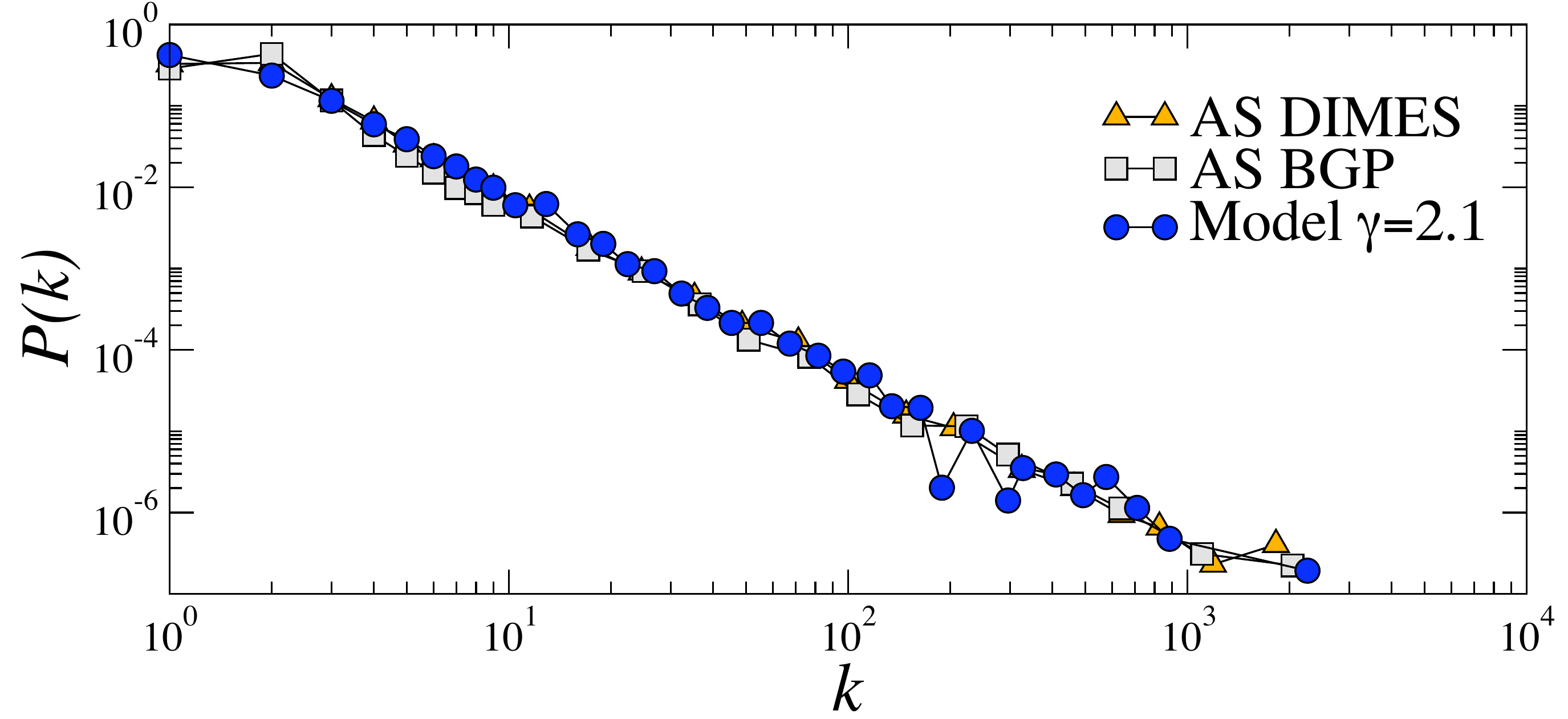}\\
\includegraphics[width=8.5cm]{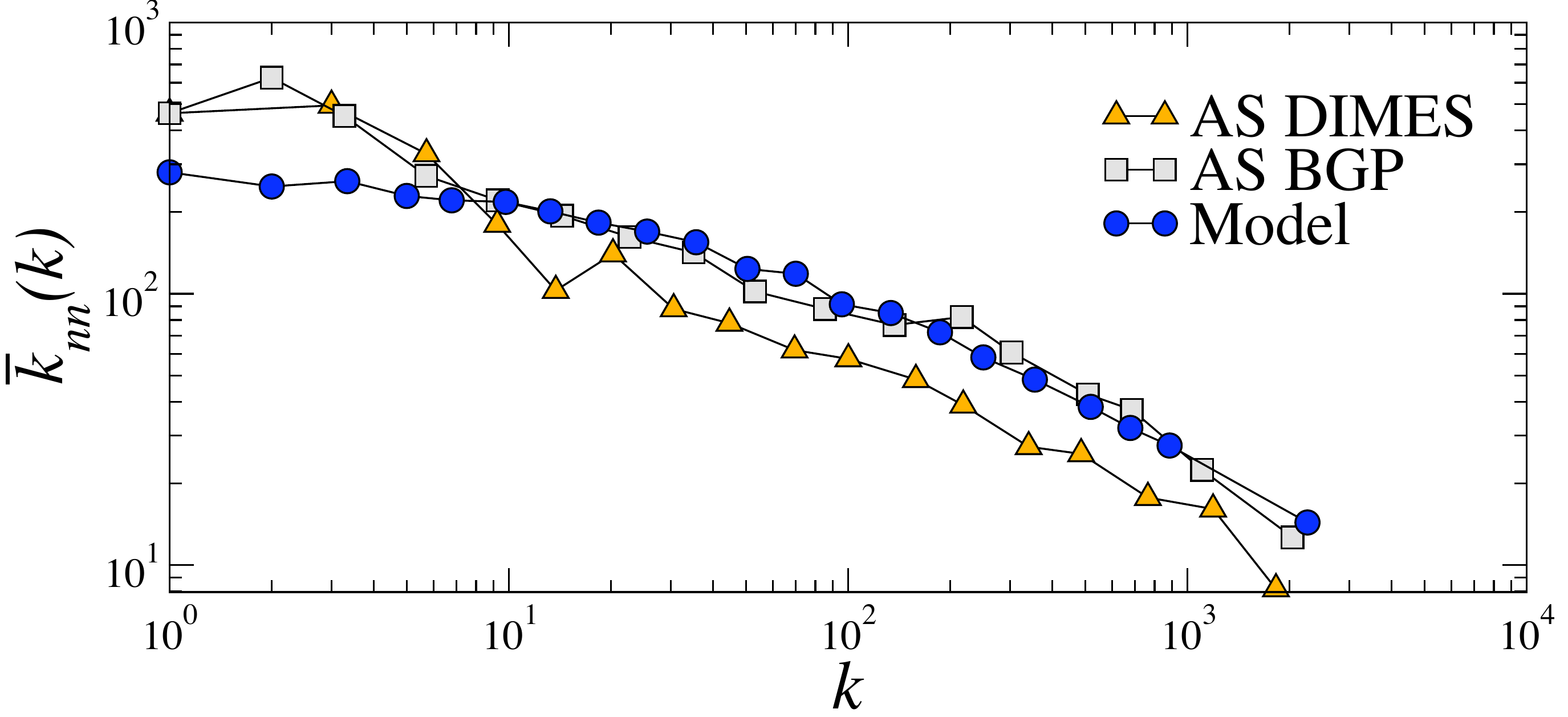}\\
\includegraphics[width=8.5cm]{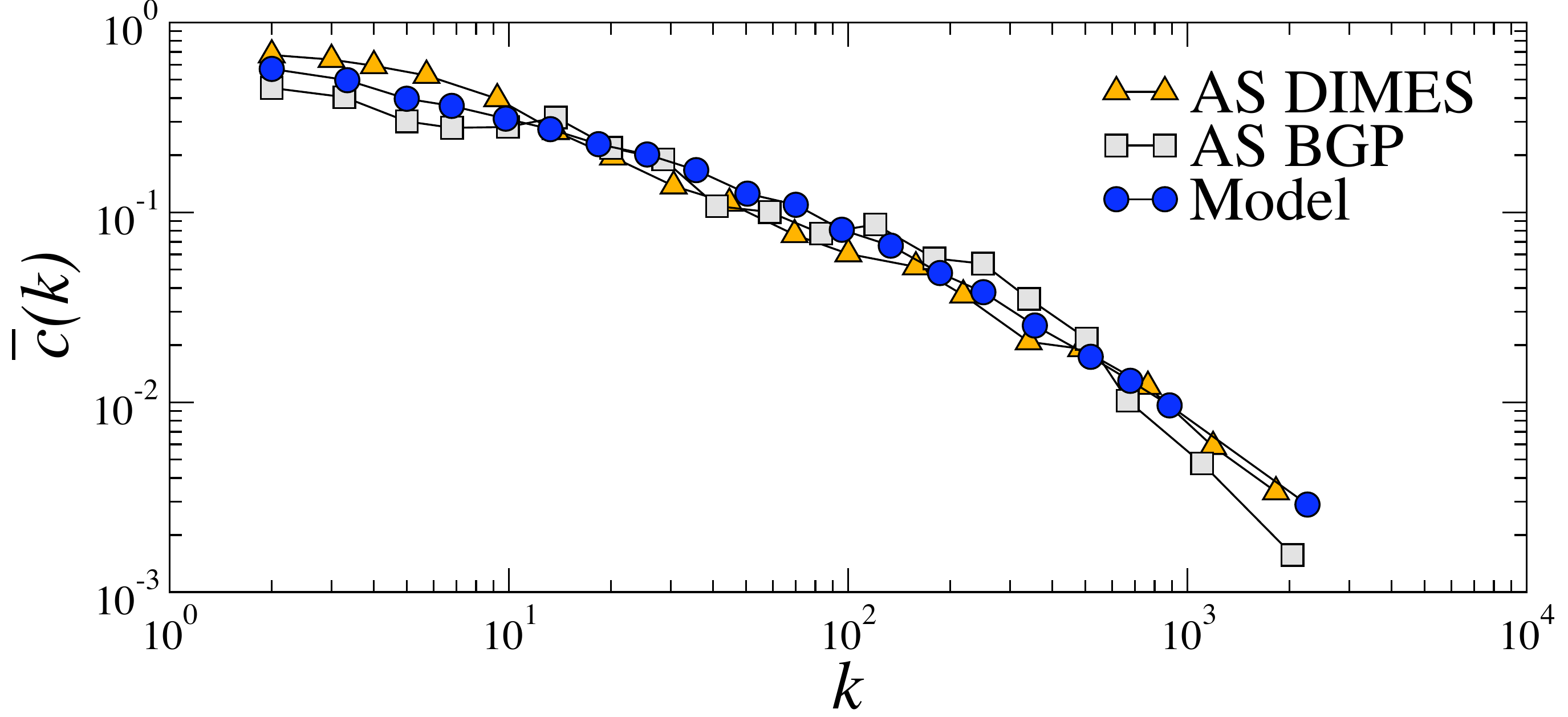}
\end{tabular}
\end{center}
\caption{Degree distribution $P(k)$, average nearest neighbours'
degree $\bar{k}_{nn}(k)$, and degree-dependent clustering
coefficient $\bar{c}(k)$ generated by our model with $\gamma=2.1$
and $\alpha=2$ compared to the same metrics for the real Internet
map as seen by BGP data and the DIMES project.} \label{Model_vs_AS}
\end{figure}

\begin{figure}[t]
\begin{center}
\begin{tabular}{c}
\includegraphics[width=8.5cm]{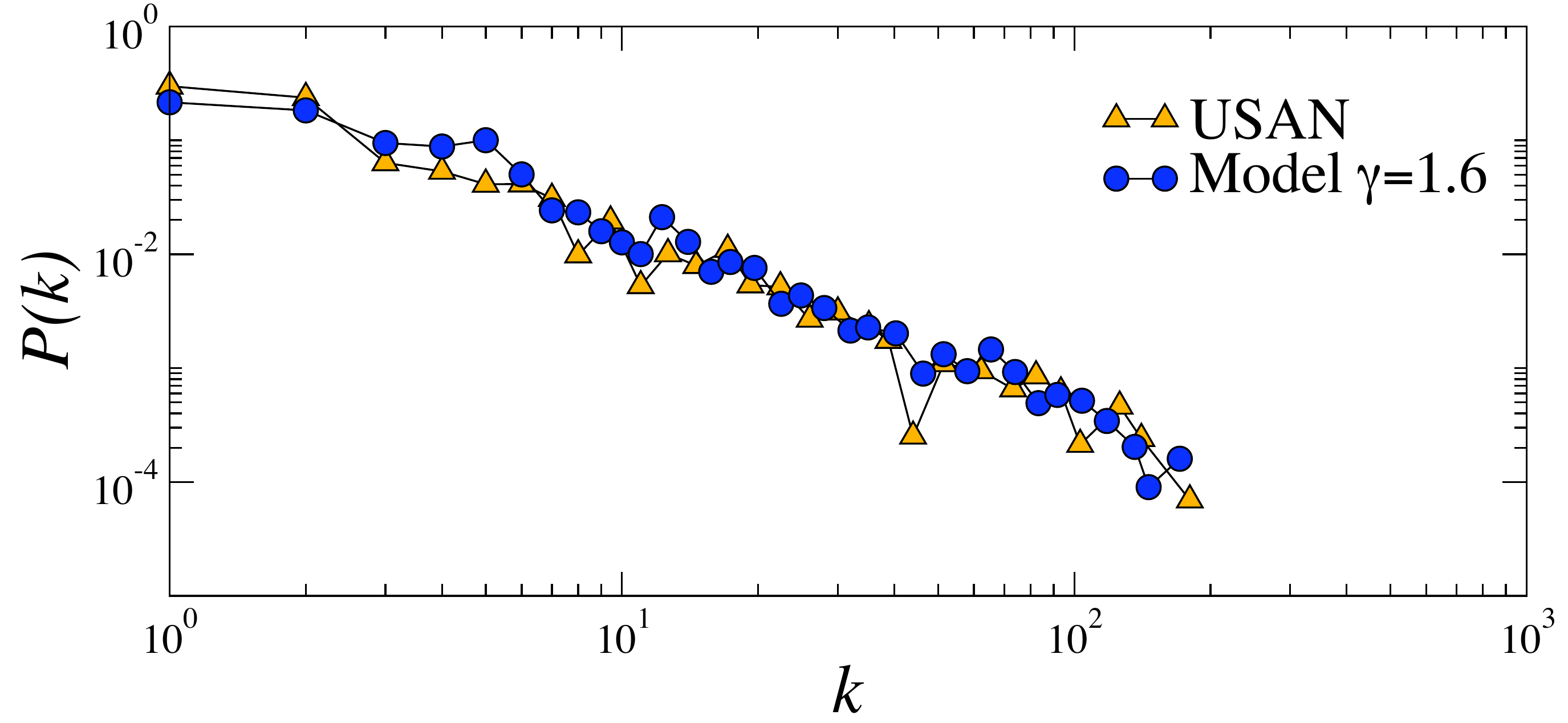}\\
\includegraphics[width=8.5cm]{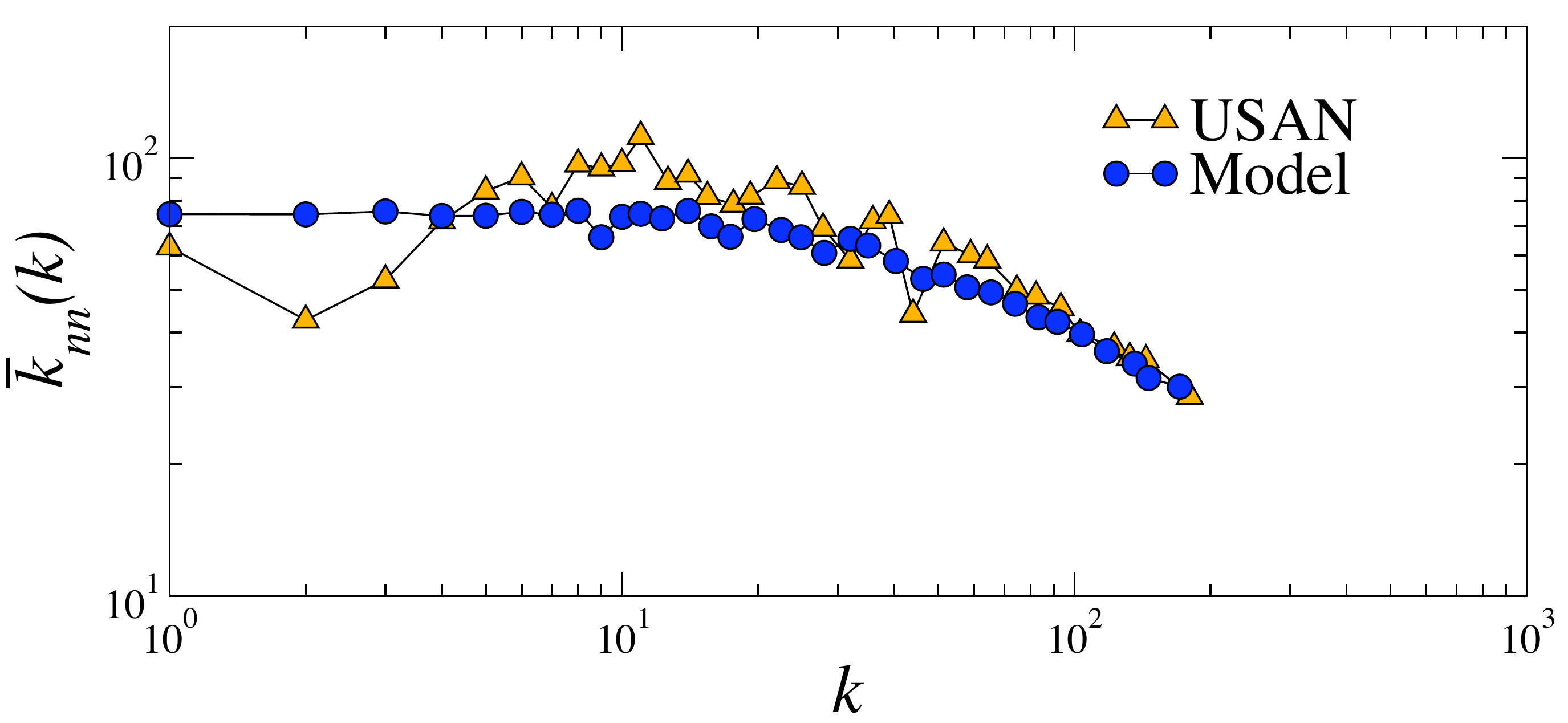}\\
\includegraphics[width=8.5cm]{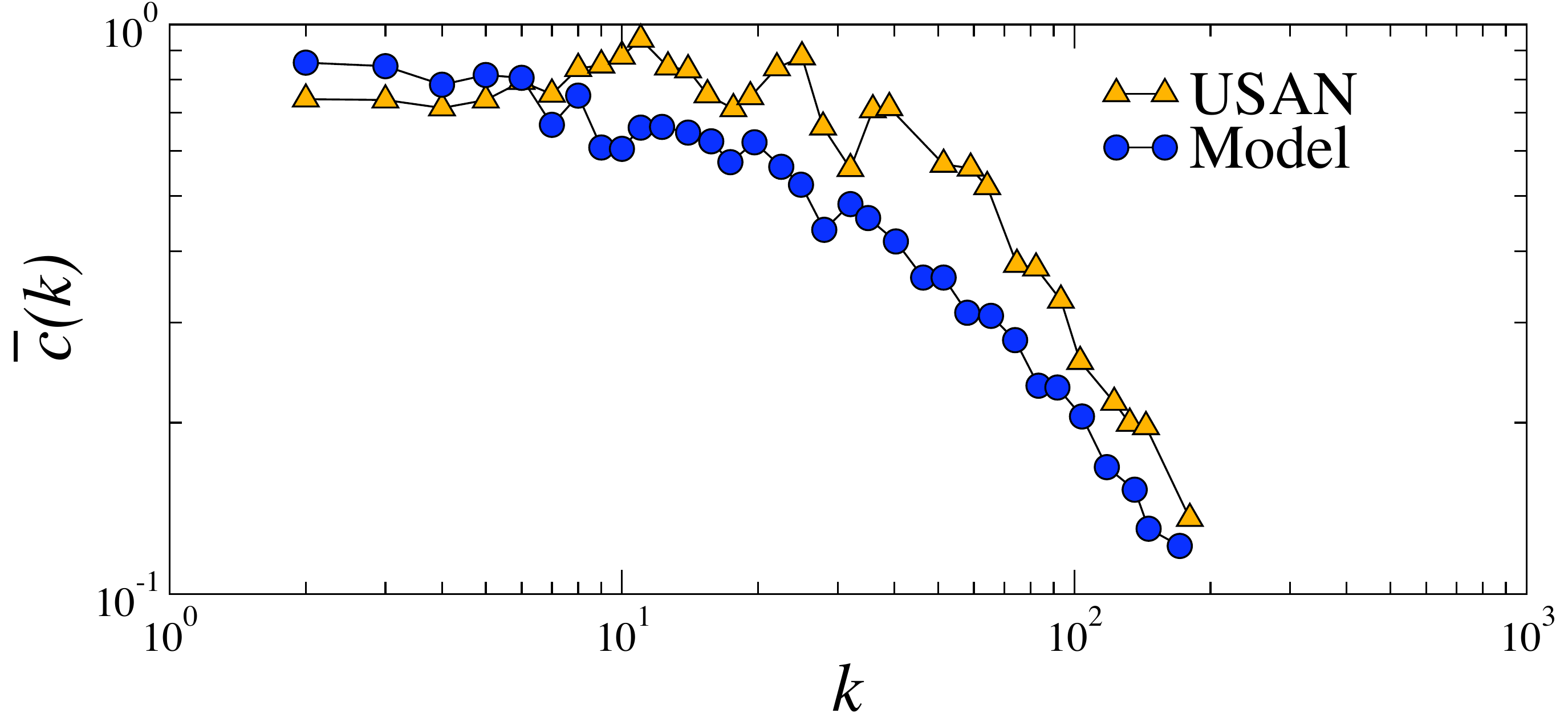}
\end{tabular}
\end{center}
\caption{Degree distribution $P(k)$, average nearest neighbours'
degree $\bar{k}_{nn}(k)$, and degree-dependent clustering
coefficient $\bar{c}(k)$ generated by our model with $\gamma=1.6$,
$\alpha=5$ and a cut-off at $k_{c}=180$ compared to the same metrics
for the real US airport network.} \label{Model_vs_USAN}
\end{figure}

Figs.~\ref{Model_vs_AS} and~\ref{Model_vs_USAN} show a comparison
of the basic topological properties of these networks with graphs
generated with the model. In the case of the AS map, we use a
truncated power law distribution $\rho(\kappa) \sim
\kappa^{-\gamma}$, $\kappa< \kappa_{c}$ with exponent $\gamma=2.1$
and $\kappa_{c}$ such that the maximum degree of the network is
$k_{c} = 2400$. For the USAN, we use $\gamma=1.6$ and a maximum
degree $k_{c}=180$, as observed in the real network. As it can be
appreciated in both figures, the matching of the model with the
empirical data is surprisingly good except for very low degree
vertices. This is particularly interesting since we are not
enforcing any mechanism to reproduce higher order statistics like
the average nearest neighbours degree $\bar{k}_{nn}(k)$ or the
degree-dependent clustering coefficient $\bar{c}(k)$. This can be
understood as a consequence of the high heterogeneity of the degree
distribution that introduces structural constraints in the
network~\cite{Park:2003,Boguna:2004}.

The airport network differs in several
ways from our modelled networks: the distribution of airports in the
geographic space is far from uniform; the airport degree
distribution does not perfectly follow a power law; and it exhibits
a sharp high-degree cut-off.
However, the structure of greedy paths is surprisingly similar to
that in our modelled networks in Fig.~\ref{fig:path_hierarchy}. The
success ratio $p_s \approx 0.64$ and average length of successful
paths $\tau \approx 2.1$ are also similar to those in our modelled
networks of the corresponding size, clustering, and degree
distribution exponent. These similarities indicate that the network
navigability characteristics depend on clustering and heterogeneity
of the airport degree distribution, and less so on how perfectly it
follows a power law.

\section{Hierarchical organization of modeled networks}

The routing process in our framework resembles guided searching for
a specific object in a complex collection of objects. Perhaps the
simplest and most general way to make a complex collection of
heterogenous objects searchable is to classify them in a
hierarchical fashion. By ``hierarchical,'' we mean that the whole
collection is split into categories (i.e., sets), sub-categories,
sub-sub-categories, and so on. Relationships between categories form
(almost) a tree, whose leaves are individual objects in the
collection~\cite{WatDoNew02,GiNe02,ClMo08,KrPa08}. Finding an object
reduces to the simpler task of navigating this tree.


\begin{figure*}[t]
\begin{center}
\includegraphics[width=17cm]{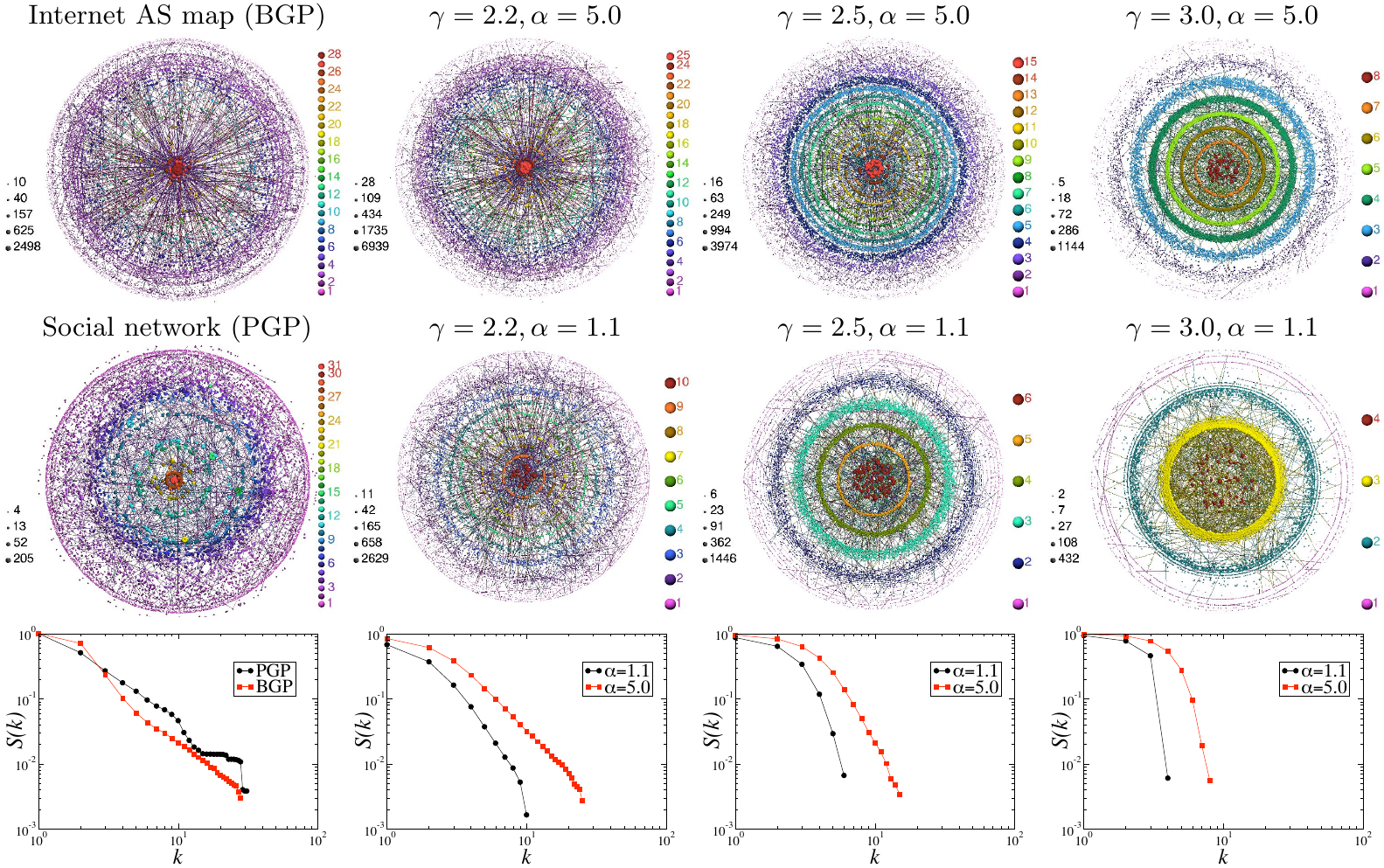}
\end{center}
\vspace{-0.5cm} \caption{{\bf $k$-core decompositions of real and
modeled networks}. The first two rows show LaNet-vi \cite{Lanet-vi}
network visualizations. All nodes are color-coded based on their
coreness (right legends) and size-coded based on their degrees (left
legends). Higher-coreness nodes are closer to circle centers. The third row shows the $k$-core
spectrum, i.e., the distribution ${\cal S}(k)$ of sizes of node sets
with coreness $k$. The first column depicts two real networks: the
AS-level Internet as seen by the Border Gateway Protocol (BGP) in
\cite{MaKrFo06} and the Pretty Good Privacy (PGP) social network
from \cite{BoPa04a}. The rest of the columns show modeled networks
for different values of power-law exponent $\gamma$ in cases with
weak ($\alpha=1.1$) and strong ($\alpha=5.0$) clustering. The
network size $N$ for all real and modeled cases is approximately
$10^4$. Similarity between real networks and modeled networks with
low $\gamma$ and high $\alpha$ is remarkable.} \label{kcore}
\end{figure*}

$k$-core decomposition \cite{bollobas98,DoGo06} is possibly the most
suitable generic tool to expose hierarchy within our modeled networks.
The $k$-core of a network is its maximal subgraph such that
all the nodes in the subgraph have $k$ or more connections to other
nodes in the subgraph. A node's coreness is the maximum $k$ such that
the $k$-core contains the node but the $k+1$-core does not.
The $k$-core structure of a network
is a form of hierarchy since a
$k+1$-core is a subset of a $k$-core. One can estimate the
quality of this hierarchy using properties of the $k$-core
spectrum, i.e., the distribution of $k$-core sizes. If the
maximum node coreness is large and if there is a rich collection of
comparably-sized $k$-cores with a wide spectrum of $k$'s, then this
hierarchy is deep and well-developed, making it potentially more navigable.
It is poor, non-navigable otherwise.

In Fig.~\ref{kcore} we feed real and modeled networks to the Large
Network visualization tool (LaNet-vi) \cite{Lanet-vi} which utilizes
node coreness to visualize the network. Fig.~\ref{kcore} shows that
networks with stronger clustering and smaller exponents of degree
distribution possess stronger $k$-core hierarchies. These
hierarchies are directly related to how networks are constructed in
our model, since nodes with higher $\kappa$ and, consequently,
higher degrees have generally higher coreness, as we can partially
see in Fig.~\ref{kcore}.

\section{The one-hop propagator of greedy routing}
\label{sec:app:propagator}

To derive the greedy-routing propagator in this appendix, we adopt a
slightly more general formalism than in the main text. Specifically,
we assume that nodes live in a generic metric space $\cal H$ and, at
the same time, have intrinsic attributes unrelated to $\cal H$.
Contrary to normed spaces or Riemannian manifolds, generic metric
spaces do not admit any coordinates, but we still use the
coordinate-based notations here to simplify the exposition below,
and denote by ${\bf x}$ nodes' coordinates in $\cal H$ and by
$\omega$ all their other, non-geometric attributes, such as their
expected degree $\kappa$. In other words, hidden variables ${\bf x}$
and $\omega$ in this general formalism represent some collections of
nodes' geometric and non-geometric hidden attributes, not just a
pair of scalar quantities. Therefore, integrations over ${\bf x}$
and $\omega$ in what follows stand merely to denote an appropriate
form of summation in each concrete case.

As in the main text, we assume that ${\bf x}$ and $\omega$ are
independent random variables so that the probability density to find
a node with hidden variables $({\bf x},\omega)$ is
\begin{equation}
\rho({\bf x},\omega)=\delta({\bf x}) \rho(\omega)/N,
\end{equation}
where $\rho(\omega)$ is the probability density of the $\omega$
variables and $\delta({\bf x})$ is the concentration of nodes in
${\cal H}$. The total number of nodes is
\begin{equation}
N=\int_{\cal H} \delta({\bf x}) d{\bf x},
\end{equation}
and the connection probability between two nodes is an integrable
decreasing function of the hidden distance between them,
\begin{equation}
r({\bf x},\omega;{\bf x}',\omega')=r[d({\bf x},{\bf
x}')/d_c(\omega,\omega')],
\end{equation}
where $d_c(\omega,\omega')$ a characteristic distance scale that
depends on $\omega$ and $\omega'$.

We define the one-step propagator of greedy routing as the
probability $G({\bf x}',\omega'|{\bf x},\omega;{\bf x}_t)$ that the
next hop after a node with hidden variables $({\bf x},\omega)$ is a
node with hidden variables $({\bf x}',\omega')$, given that the
final destination is located at ${\bf x}_t$.

To further simplify the notations below, we label the set of
variables $({\bf x},\omega)$ as a generic hidden variable $h$ and
undo this notation change at the end of the calculations according
to the following rules:
\begin{equation}
\begin{array}{rcl}
({\bf x},\omega) & \longrightarrow & h\\
\rho({\bf x},\omega) & \longrightarrow & \rho(h)\\
d{\bf x} d\omega & \longrightarrow & dh\\
r({\bf x},\omega;{\bf x}',\omega')& \longrightarrow & r(h,h').
\end{array}
\end{equation}

We begin the propagator derivation assuming that a particular
network instance has a configuration given by
$\{h,h_{t},h_{1},\cdots,h_{N-2}\} \equiv \{h,h_{t};\{h_{j}\}\}$ with
$j=1,\cdots,N-2$, where $h$ and $h_t$ denote the hidden variables of
the current hop and the destination, respectively. In this
particular network configuration, the probability that the current
node's next hop is a particular node $i$ with hidden variable
$h_{i}$ is the probability that the current node is connected to $i$
but disconnected to all nodes that are closer to the destination
than $i$,
\begin{widetext}
\begin{equation}
\mbox{Prob}(i |h,h_t;\{h_{j}\})
=r(h,h_i) \displaystyle{\prod_{j (\ne i)=1}^{N-2}}
\left[1-r(h,h_j)\right]^{\Theta
\left[d(h_i,h_t)-d(h_j,h_t)\right]},
\end{equation}
\end{widetext}
where $\Theta(\cdot)$ is the Heaviside step function. Taking the
average over all possible configurations
$\{h_1,\cdots,h_{i-1},h_{i+1},\cdots,h_{N-2}\}$ excluding node $i$,
we obtain
\begin{equation}
\mbox{Prob}(i|h,h_t;h_i)=r(h,h_i)\left(1-\frac{1}{N-3}
\bar{k}(h|h_i,h_t)\right)^{N-3},
\end{equation}
where
\begin{equation}
\bar{k}(h|h_i,h_t)=(N-3)\int_{d(h_i,h_t)<d(h',h_t)} \rho(h')
r(h,h') dh'
\end{equation}
is the average number of connections between the current node and
nodes closer to the destination than node $i$, excluding $i$ and
$t$.

The probability that the next hop has hidden variable $h'$,
regardless of its label, i.e., index $i$, is
\begin{equation}
\mbox{Prob}(h'|h,h_t)=\sum_{i=1}^{N-2} \rho(h') \mbox{Prob}(i|h,h_t;h').
\end{equation}
In the case of sparse networks, $\bar{k}(h|h',h_t)$ is a finite
quantity. Taking the limit of large $N$, the above expression
simplifies to
\begin{equation}
\mbox{Prob}(h'|h,h_t)=N \rho(h') r(h,h') e^{-\bar{k}(h|h',h_t)}.
\end{equation}
Yet, this equation is not a properly normalized probability density
function for the variable $h'$ since node $h$ can have degree zero
with some probability. If we consider only nodes with degrees
greater than zero, then the normalization factor is given by
$1-e^{-\bar{k}(h)}$. Therefore, the properly normalized propagator
is finally
\begin{equation}
G(h'|h,h_t)=\displaystyle{\frac{N \rho(h') r(h,h') e^{-\bar{k}(h|h',h_t)}}{1-e^{-\bar{k}(h)}}}.
\end{equation}

We now undo the notation change and express this propagator in terms
of our mixed coordinates:
\begin{widetext}
\begin{equation}
G({\bf x}',\omega'|{\bf x},\omega;{\bf x}_t)=\frac{\delta({\bf x}')\rho(\omega')}{1-e^{-\bar{k}({\bf x},\omega)}}
r\left[\frac{d({\bf x},{\bf x}')}{d_{c}(\omega,\omega')}\right]
e^{-\bar{k}({\bf x},\omega|{\bf x}',{\bf x}_t)},
\label{prop}
\end{equation}
with
\begin{equation} \label{kmeandistance}
\bar{k}({\bf x},\omega|{\bf x}',{\bf x}_t) = \int_{d({\bf x}',{\bf
x}_{t})>d({\bf y},{\bf x}_{t})} d{\bf y}\int d\omega' \delta({\bf
y})\rho(\omega') r\left[\frac{d({\bf x},{\bf
y})}{d_{c}(\omega,\omega')}\right].
\end{equation}
\end{widetext}

In the particular case of the $\mathbb{S}^1$ model, we can express
this propagator in terms of relative hidden distances instead of
absolute coordinates. Namely, $G(d',\omega'|d,\omega)$ is the
probability that an $\omega$-labeled node, e.g., a node with
expected degree $\kappa=\omega$, at hidden distance $d$ from the
destination has as the next hop an $\omega'$-labeled node at hidden
distance $d'$ from the destination. After tedious calculations, the
resulting expression reads:
\begin{widetext}
\begin{equation}
\footnotesize
\hspace{-0.cm}
G(d',\omega'|d,\omega)=
\left\{
\begin{array}{ll}
\frac{(\gamma-1)}{\omega'^{\gamma}}
\left[\frac{1}{(1+\frac{d-d'}{\mu \omega \omega'})^{\alpha}}+\frac{1}{(1+\frac{d+d'}{\mu \omega \omega'})^{\alpha}}
\right]\mbox{exp}
\left\{
\frac{(1-\gamma)\mu \omega}{\alpha-1}\left[ {\cal B}(\frac{d-d'}{\mu \omega},\gamma-2,2-\alpha)-{\cal B}(\frac{d+d'}{\mu \omega},\gamma-2,2-\alpha)\right]
\right\}& ;d'\le d
\\[0.5cm]
\frac{(\gamma-1)}{\omega'^{\gamma}}
\left[ \frac{1}{(1+\frac{d'-d}{\mu \omega \omega'})^{\alpha}}+\frac{1}{(1+\frac{d+d'}{\mu \omega \omega'})^{\alpha}}
\right]\mbox{exp}
\left\{
\frac{(1-\gamma)\mu \omega}{\alpha-1}\left[\frac{2}{\gamma-2} -{\cal B}(\frac{d'-d}{\mu \omega},\gamma-2,2-\alpha)-{\cal B}(\frac{d+d'}{\mu \omega},\gamma-2,2-\alpha)\right]
\right\}& ;d'>d
\end{array}
\right.,
\end{equation}
\end{widetext}
where we have defined function
\begin{equation}
{\cal B}(z,a,b) \equiv z^{-a}\int_0^z t^{a-1} (1+t)^{b-1} dt,
\end{equation}
which is somewhat similar to the incomplete beta function
$B(z,a,b)=\int_0^z t^{a-1} (1-t)^{b-1} dt$.

\begin{figure}[t]
\begin{center}
\vspace{-0cm}
\includegraphics[width=8.5cm]{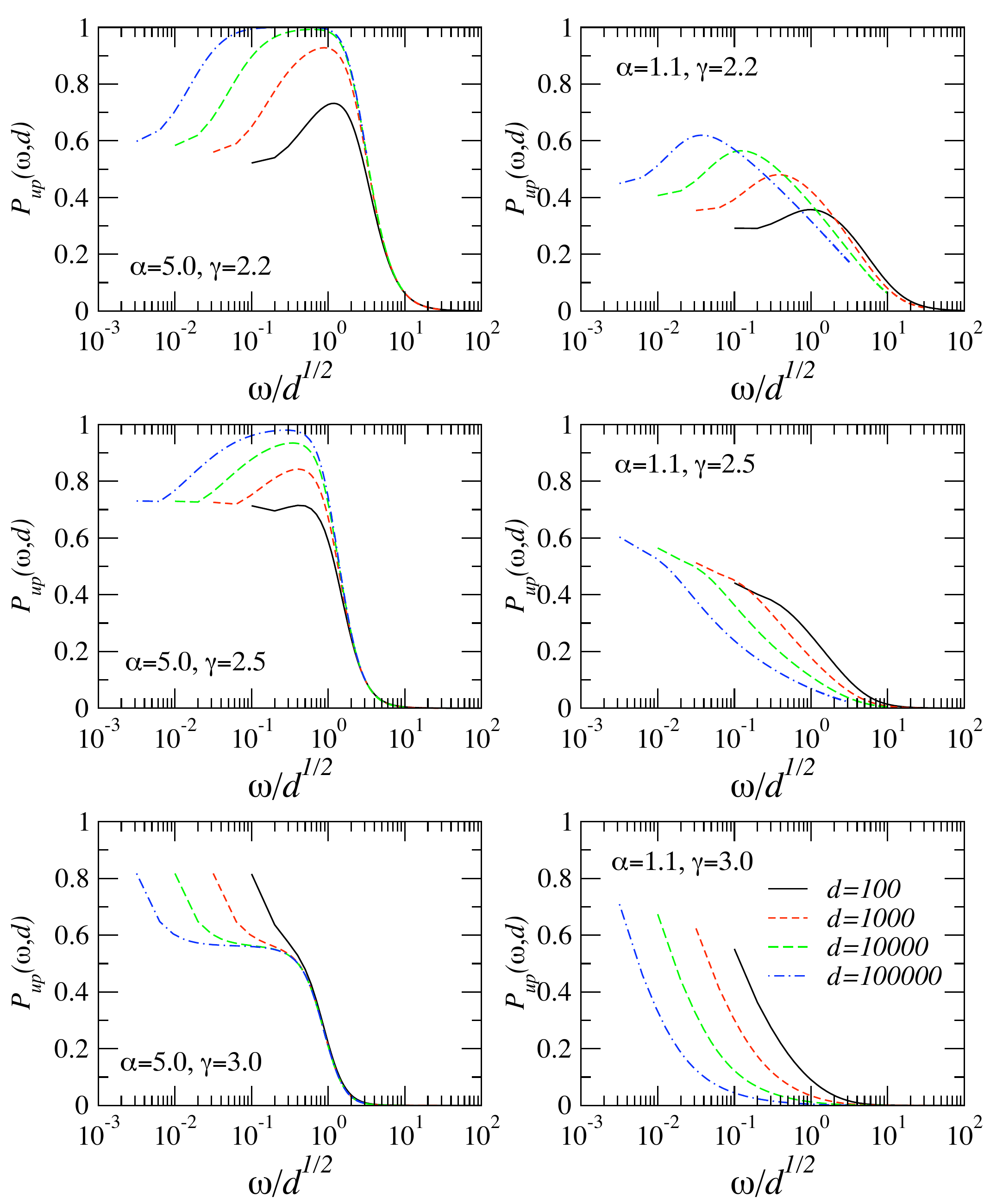}
\end{center}
\vspace{-0.5cm} \caption{Probability $P_{up}(\omega/d^{1/2},d)$.}
\label{fig:P_up-proper-scaling}
\end{figure}

One of the informative quantities elucidating the structure of
greedy-routing paths is the probability $P_{up}(\omega,d)$ that the
next hop after an $\omega$-labeled node at distance $d$ from the
destination has a higher value of $\omega$. The greedy-routing
propagator defines this probability as
\begin{equation}
P_{up}(\omega,d) = \int_{\omega' \geq \omega} d\omega' \int_{d'<d}
dd' G(d',\omega'|d,\omega),
\end{equation}
and we show $P_{up}(\omega/d^{1/2},d)$ in
Fig.~\ref{fig:P_up-proper-scaling}. We see that the proper scaling
of $\omega_c \sim d^{1/2}$, where $\omega_c$ is the critical value
of $\omega$ above which $P_{up}(\omega,d)$ quickly drops to zero, is
present only when clustering is strong. Furthermore,
$P_{up}(\omega,d)$ is an increasing function of $\omega$ for small
$\omega$'s only when the degree distribution exponent $\gamma$ is
close to $2$. A combination of these two effects guarantees that the
layout of greedy routes properly adapts to increasing distances or
graph sizes, thus making networks with strong clustering and
$\gamma$'s greater than but close to $2$ navigable.


\begin{thebibliography}{48}
\expandafter\ifx\csname natexlab\endcsname\relax\def\natexlab#1{#1}\fi
\expandafter\ifx\csname bibnamefont\endcsname\relax
  \def\bibnamefont#1{#1}\fi
\expandafter\ifx\csname bibfnamefont\endcsname\relax
  \def\bibfnamefont#1{#1}\fi
\expandafter\ifx\csname citenamefont\endcsname\relax
  \def\citenamefont#1{#1}\fi
\expandafter\ifx\csname url\endcsname\relax
  \def\url#1{\texttt{#1}}\fi
\expandafter\ifx\csname urlprefix\endcsname\relax\def\urlprefix{URL }\fi
\providecommand{\bibinfo}[2]{#2}
\providecommand{\eprint}[2][]{\url{#2}}

\bibitem[{\citenamefont{Albert and Barab{\'a}si}(2002)}]{Albert:2002}
\bibinfo{author}{\bibfnamefont{R.}~\bibnamefont{Albert}} \bibnamefont{and}
  \bibinfo{author}{\bibfnamefont{A.-L.} \bibnamefont{Barab{\'a}si}},
  \bibinfo{journal}{Rev. Mod. Phys.} \textbf{\bibinfo{volume}{74}},
  \bibinfo{pages}{47} (\bibinfo{year}{2002}).

\bibitem[{\citenamefont{Newman}(2003)}]{newman03c-review}
\bibinfo{author}{\bibfnamefont{M.~E.~J.} \bibnamefont{Newman}},
  \bibinfo{journal}{SIAM Rev} \textbf{\bibinfo{volume}{45}},
  \bibinfo{pages}{167} (\bibinfo{year}{2003}).

\bibitem[{\citenamefont{Dorogovtsev and Mendes}(2003)}]{Dorogovtsev:2003}
\bibinfo{author}{\bibfnamefont{S.~N.} \bibnamefont{Dorogovtsev}}
  \bibnamefont{and} \bibinfo{author}{\bibfnamefont{J.~F.~F.}
  \bibnamefont{Mendes}}, \emph{\bibinfo{title}{Evolution of networks: From
  biological nets to the Internet and WWW}} (\bibinfo{publisher}{Oxford
  University Press}, \bibinfo{address}{Oxford}, \bibinfo{year}{2003}).

\bibitem[{\citenamefont{Boccaletti et~al.}(2006)\citenamefont{Boccaletti,
  Latora, Moreno, Chavez, and Hwang}}]{Boccaletti:2006}
\bibinfo{author}{\bibfnamefont{S.}~\bibnamefont{Boccaletti}},
  \bibinfo{author}{\bibfnamefont{V.}~\bibnamefont{Latora}},
  \bibinfo{author}{\bibfnamefont{Y.}~\bibnamefont{Moreno}},
  \bibinfo{author}{\bibfnamefont{M.}~\bibnamefont{Chavez}}, \bibnamefont{and}
  \bibinfo{author}{\bibfnamefont{D.-U.} \bibnamefont{Hwang}},
  \bibinfo{journal}{Phys. Rep.} \textbf{\bibinfo{volume}{424}},
  \bibinfo{pages}{175} (\bibinfo{year}{2006}).

\bibitem[{\citenamefont{Castells}(1996)}]{Castells:2006}
\bibinfo{author}{\bibfnamefont{M.}~\bibnamefont{Castells}},
  \emph{\bibinfo{title}{The rise of the network society}}
  (\bibinfo{publisher}{Blackwell Publishing, Oxford}, \bibinfo{year}{1996}).

\bibitem[{\citenamefont{Pastor-Satorras and
  Vespignani}(2004)}]{RomusVespasbook}
\bibinfo{author}{\bibfnamefont{R.}~\bibnamefont{Pastor-Satorras}}
  \bibnamefont{and}
  \bibinfo{author}{\bibfnamefont{A.}~\bibnamefont{Vespignani}},
  \emph{\bibinfo{title}{Evolution and Structure of the Internet. A Statistical
  Physics Approach}} (\bibinfo{publisher}{Cambridge University Press},
  \bibinfo{address}{Cambridge}, \bibinfo{year}{2004}).

\bibitem[{\citenamefont{Watts et~al.}(2002)\citenamefont{Watts, Dodds, and
  Newman}}]{WatDoNew02}
\bibinfo{author}{\bibfnamefont{D.~J.} \bibnamefont{Watts}},
  \bibinfo{author}{\bibfnamefont{P.~S.} \bibnamefont{Dodds}}, \bibnamefont{and}
  \bibinfo{author}{\bibfnamefont{M.~E.~J.} \bibnamefont{Newman}},
  \bibinfo{journal}{Science} \textbf{\bibinfo{volume}{296}},
  \bibinfo{pages}{1302} (\bibinfo{year}{2002}).

\bibitem[{\citenamefont{Girvan and Newman}(2002)}]{GiNe02}
\bibinfo{author}{\bibfnamefont{M.}~\bibnamefont{Girvan}} \bibnamefont{and}
  \bibinfo{author}{\bibfnamefont{M.~E.~J.} \bibnamefont{Newman}},
  \bibinfo{journal}{Proc. Nat. Acad. Sci. USA} \textbf{\bibinfo{volume}{99}},
  \bibinfo{pages}{7821} (\bibinfo{year}{2002}).

\bibitem[{\citenamefont{Menczer}(2002)}]{menczer02-pnas}
\bibinfo{author}{\bibfnamefont{F.}~\bibnamefont{Menczer}},
  \bibinfo{journal}{Proc. Nat. Acad. Sci. USA} \textbf{\bibinfo{volume}{99}},
  \bibinfo{pages}{14014} (\bibinfo{year}{2002}).

\bibitem[{\citenamefont{Leicht et~al.}(2006)\citenamefont{Leicht, Holme, and
  Newman}}]{LeHoNe06}
\bibinfo{author}{\bibfnamefont{E.~A.} \bibnamefont{Leicht}},
  \bibinfo{author}{\bibfnamefont{P.}~\bibnamefont{Holme}}, \bibnamefont{and}
  \bibinfo{author}{\bibfnamefont{M.~E.~J.} \bibnamefont{Newman}},
  \bibinfo{journal}{Phys Rev E} \textbf{\bibinfo{volume}{73}},
  \bibinfo{pages}{026120} (\bibinfo{year}{2006}).

\bibitem[{\citenamefont{Crandall et~al.}(2008)\citenamefont{Crandall, Cosley,
  Huttenlocher, Kleinberg, and Suri}}]{CraCo08}
\bibinfo{author}{\bibfnamefont{D.}~\bibnamefont{Crandall}},
  \bibinfo{author}{\bibfnamefont{D.}~\bibnamefont{Cosley}},
  \bibinfo{author}{\bibfnamefont{D.}~\bibnamefont{Huttenlocher}},
  \bibinfo{author}{\bibfnamefont{J.}~\bibnamefont{Kleinberg}},
  \bibnamefont{and} \bibinfo{author}{\bibfnamefont{S.}~\bibnamefont{Suri}}, in
  \emph{\bibinfo{booktitle}{Proceedings of the ACM International Conference on
  Knowledge Discovery and Data Mining}} (\bibinfo{publisher}{ACM},
  \bibinfo{year}{2008}).

\bibitem[{\citenamefont{Clauset et~al.}(2008)\citenamefont{Clauset, Moore, and
  Newman}}]{ClMo08}
\bibinfo{author}{\bibfnamefont{A.}~\bibnamefont{Clauset}},
  \bibinfo{author}{\bibfnamefont{C.}~\bibnamefont{Moore}}, \bibnamefont{and}
  \bibinfo{author}{\bibfnamefont{M.~E.~J.} \bibnamefont{Newman}},
  \bibinfo{journal}{Nature} \textbf{\bibinfo{volume}{453}}, \bibinfo{pages}{98}
  (\bibinfo{year}{2008}).

\bibitem[{\citenamefont{\'{A}ngeles Serrano
  et~al.}(2008)\citenamefont{\'{A}ngeles Serrano, Krioukov, and
  Bogu{\~{n}}\'{a}}}]{SerKriBog07}
\bibinfo{author}{\bibfnamefont{M.}~\bibnamefont{\'{A}ngeles Serrano}},
  \bibinfo{author}{\bibfnamefont{D.}~\bibnamefont{Krioukov}}, \bibnamefont{and}
  \bibinfo{author}{\bibfnamefont{M.}~\bibnamefont{Bogu{\~{n}}\'{a}}},
  \bibinfo{journal}{Phys Rev Lett} \textbf{\bibinfo{volume}{100}},
  \bibinfo{pages}{078701} (\bibinfo{year}{2008}).

\bibitem[{\citenamefont{Travers and Milgram}(1969)}]{TraMi69}
\bibinfo{author}{\bibfnamefont{J.}~\bibnamefont{Travers}} \bibnamefont{and}
  \bibinfo{author}{\bibfnamefont{S.}~\bibnamefont{Milgram}},
  \bibinfo{journal}{Sociometry} \textbf{\bibinfo{volume}{32}},
  \bibinfo{pages}{425} (\bibinfo{year}{1969}).

\bibitem[{\citenamefont{Watts and Strogatz}(1998)}]{Watts:1998}
\bibinfo{author}{\bibfnamefont{D.~J.} \bibnamefont{Watts}} \bibnamefont{and}
  \bibinfo{author}{\bibfnamefont{S.~H.} \bibnamefont{Strogatz}},
  \bibinfo{journal}{Nature} \textbf{\bibinfo{volume}{393}},
  \bibinfo{pages}{440} (\bibinfo{year}{1998}).

\bibitem[{\citenamefont{Kleinberg}(2000{\natexlab{a}})}]{kleinberg00-nature}
\bibinfo{author}{\bibfnamefont{J.}~\bibnamefont{Kleinberg}},
  \bibinfo{journal}{Nature} \textbf{\bibinfo{volume}{406}},
  \bibinfo{pages}{845} (\bibinfo{year}{2000}{\natexlab{a}}).

\bibitem[{\citenamefont{Kleinberg}(2000{\natexlab{b}})}]{kleinberg00-stoc}
\bibinfo{author}{\bibfnamefont{J.}~\bibnamefont{Kleinberg}}, in
  \emph{\bibinfo{booktitle}{STOC '00: Proceedings of the thirty-second annual
  ACM symposium on Theory of computing}} (\bibinfo{publisher}{ACM},
  \bibinfo{address}{New York, NY, USA}, \bibinfo{year}{2000}{\natexlab{b}}),
  pp. \bibinfo{pages}{163--170}, ISBN \bibinfo{isbn}{1-58113-184-4}.

\bibitem[{\citenamefont{Kleinberg}(2001)}]{kleinberg01}
\bibinfo{author}{\bibfnamefont{J.~M.} \bibnamefont{Kleinberg}}, in
  \emph{\bibinfo{booktitle}{NIPS}}, edited by
  \bibinfo{editor}{\bibfnamefont{T.~G.} \bibnamefont{Dietterich}},
  \bibinfo{editor}{\bibfnamefont{S.}~\bibnamefont{Becker}}, \bibnamefont{and}
  \bibinfo{editor}{\bibfnamefont{Z.}~\bibnamefont{Ghahramani}}
  (\bibinfo{publisher}{MIT Press}, \bibinfo{year}{2001}), pp.
  \bibinfo{pages}{431--438}.

\bibitem[{\citenamefont{Manku et~al.}(2004)\citenamefont{Manku, Naor, and
  Wieder}}]{MaNaWe04}
\bibinfo{author}{\bibfnamefont{G.~S.} \bibnamefont{Manku}},
  \bibinfo{author}{\bibfnamefont{M.}~\bibnamefont{Naor}}, \bibnamefont{and}
  \bibinfo{author}{\bibfnamefont{U.}~\bibnamefont{Wieder}}, in
  \emph{\bibinfo{booktitle}{Proceedings of the 36th ACM Symposium on Theory of
  Computing (STOC)}} (\bibinfo{publisher}{ACM}, \bibinfo{year}{2004}), pp.
  \bibinfo{pages}{54--63}.

\bibitem[{\citenamefont{Martel}(2004)}]{MaNg04}
\bibinfo{author}{\bibfnamefont{C.}~\bibnamefont{Martel}}, in
  \emph{\bibinfo{booktitle}{23rd ACM Symp. on Principles of Distributed
  Computing (PODC)}} (\bibinfo{publisher}{ACM Press}, \bibinfo{year}{2004}),
  pp. \bibinfo{pages}{179--188}.

\bibitem[{\citenamefont{Nguyen and Martel}(2005)}]{NgMa05}
\bibinfo{author}{\bibfnamefont{V.}~\bibnamefont{Nguyen}} \bibnamefont{and}
  \bibinfo{author}{\bibfnamefont{C.}~\bibnamefont{Martel}}, in
  \emph{\bibinfo{booktitle}{SODA '05: Proceedings of the sixteenth annual
  ACM-SIAM symposium on Discrete algorithms}} (\bibinfo{publisher}{Society for
  Industrial and Applied Mathematics}, \bibinfo{address}{Philadelphia, PA,
  USA}, \bibinfo{year}{2005}), pp. \bibinfo{pages}{311--320}, ISBN
  \bibinfo{isbn}{0-89871-585-7}.

\bibitem[{\citenamefont{Nguyen and Martel}(2008)}]{NgMa08}
\bibinfo{author}{\bibfnamefont{V.}~\bibnamefont{Nguyen}} \bibnamefont{and}
  \bibinfo{author}{\bibfnamefont{C.}~\bibnamefont{Martel}}, in
  \emph{\bibinfo{booktitle}{Proceedings of the Fourth Workshop on Analytic
  Algorithmics and Combinatorics}} (\bibinfo{publisher}{SIAM},
  \bibinfo{year}{2008}), pp. \bibinfo{pages}{213--227}.

\bibitem[{\citenamefont{Simsek and Jensen}(2005)}]{SiJe05}
\bibinfo{author}{\bibfnamefont{{\"O}.}~\bibnamefont{Simsek}} \bibnamefont{and}
  \bibinfo{author}{\bibfnamefont{D.}~\bibnamefont{Jensen}}, in
  \emph{\bibinfo{booktitle}{IJCAI}}, edited by
  \bibinfo{editor}{\bibfnamefont{L.~P.} \bibnamefont{Kaelbling}}
  \bibnamefont{and} \bibinfo{editor}{\bibfnamefont{A.}~\bibnamefont{Saffiotti}}
  (\bibinfo{publisher}{Professional Book Center}, \bibinfo{year}{2005}), pp.
  \bibinfo{pages}{304--310}, ISBN \bibinfo{isbn}{0938075934}.

\bibitem[{\citenamefont{Lebhar and Schabanel}(2004)}]{LeSha04}
\bibinfo{author}{\bibfnamefont{E.}~\bibnamefont{Lebhar}} \bibnamefont{and}
  \bibinfo{author}{\bibfnamefont{N.}~\bibnamefont{Schabanel}}, in
  \emph{\bibinfo{booktitle}{ICALP}}, edited by
  \bibinfo{editor}{\bibfnamefont{J.}~\bibnamefont{D\'{\i}az}},
  \bibinfo{editor}{\bibfnamefont{J.}~\bibnamefont{Karhum{\"a}ki}},
  \bibinfo{editor}{\bibfnamefont{A.}~\bibnamefont{Lepist{\"o}}},
  \bibnamefont{and} \bibinfo{editor}{\bibfnamefont{D.}~\bibnamefont{Sannella}}
  (\bibinfo{publisher}{Springer}, \bibinfo{year}{2004}), vol.
  \bibinfo{volume}{3142} of \emph{\bibinfo{series}{Lecture Notes in Computer
  Science}}, pp. \bibinfo{pages}{894--905}, ISBN \bibinfo{isbn}{3-540-22849-7}.

\bibitem[{\citenamefont{Fraigniaud}(2005)}]{fraigniaud05}
\bibinfo{author}{\bibfnamefont{P.}~\bibnamefont{Fraigniaud}}, in
  \emph{\bibinfo{booktitle}{ESA}}, edited by
  \bibinfo{editor}{\bibfnamefont{G.~S.} \bibnamefont{Brodal}} \bibnamefont{and}
  \bibinfo{editor}{\bibfnamefont{S.}~\bibnamefont{Leonardi}}
  (\bibinfo{publisher}{Springer}, \bibinfo{year}{2005}), vol.
  \bibinfo{volume}{3669} of \emph{\bibinfo{series}{Lecture Notes in Computer
  Science}}, pp. \bibinfo{pages}{791--802}, ISBN \bibinfo{isbn}{3-540-29118-0}.

\bibitem[{\citenamefont{Fraigniaud et~al.}(2006)\citenamefont{Fraigniaud,
  Gavoille, and Paul}}]{FraGaPa06}
\bibinfo{author}{\bibfnamefont{P.}~\bibnamefont{Fraigniaud}},
  \bibinfo{author}{\bibfnamefont{C.}~\bibnamefont{Gavoille}}, \bibnamefont{and}
  \bibinfo{author}{\bibfnamefont{C.}~\bibnamefont{Paul}},
  \bibinfo{journal}{Distrib Comput} \textbf{\bibinfo{volume}{18}},
  \bibinfo{pages}{279} (\bibinfo{year}{2006}).

\bibitem[{\citenamefont{Fraigniaud et~al.}(2007)\citenamefont{Fraigniaud,
  Gavoille, Kosowski, Lebhar, and Lotker}}]{FraGa07}
\bibinfo{author}{\bibfnamefont{P.}~\bibnamefont{Fraigniaud}},
  \bibinfo{author}{\bibfnamefont{C.}~\bibnamefont{Gavoille}},
  \bibinfo{author}{\bibfnamefont{A.}~\bibnamefont{Kosowski}},
  \bibinfo{author}{\bibfnamefont{E.}~\bibnamefont{Lebhar}}, \bibnamefont{and}
  \bibinfo{author}{\bibfnamefont{Z.}~\bibnamefont{Lotker}}, in
  \emph{\bibinfo{booktitle}{SPAA}}, edited by
  \bibinfo{editor}{\bibfnamefont{P.~B.} \bibnamefont{Gibbons}}
  \bibnamefont{and}
  \bibinfo{editor}{\bibfnamefont{C.}~\bibnamefont{Scheideler}}
  (\bibinfo{publisher}{ACM}, \bibinfo{year}{2007}), pp. \bibinfo{pages}{1--7},
  ISBN \bibinfo{isbn}{978-1-59593-667-7}.

\bibitem[{\citenamefont{Fraigniaud and Gavoille}(2008)}]{FraGa08}
\bibinfo{author}{\bibfnamefont{P.}~\bibnamefont{Fraigniaud}} \bibnamefont{and}
  \bibinfo{author}{\bibfnamefont{C.}~\bibnamefont{Gavoille}}, in
  \emph{\bibinfo{booktitle}{SPAA}}, edited by
  \bibinfo{editor}{\bibfnamefont{F.~M.} \bibnamefont{auf~der Heide}}
  \bibnamefont{and} \bibinfo{editor}{\bibfnamefont{N.}~\bibnamefont{Shavit}}
  (\bibinfo{publisher}{ACM}, \bibinfo{year}{2008}), pp.
  \bibinfo{pages}{62--69}, ISBN \bibinfo{isbn}{978-1-59593-973-9}.

\bibitem[{\citenamefont{Chaintreau et~al.}(2008)\citenamefont{Chaintreau,
  Fraigniaud, and Lebhar}}]{ChaFra08}
\bibinfo{author}{\bibfnamefont{A.}~\bibnamefont{Chaintreau}},
  \bibinfo{author}{\bibfnamefont{P.}~\bibnamefont{Fraigniaud}},
  \bibnamefont{and} \bibinfo{author}{\bibfnamefont{E.}~\bibnamefont{Lebhar}},
  in \emph{\bibinfo{booktitle}{ICALP (1)}}, edited by
  \bibinfo{editor}{\bibfnamefont{L.}~\bibnamefont{Aceto}},
  \bibinfo{editor}{\bibfnamefont{I.}~\bibnamefont{Damg{\aa}rd}},
  \bibinfo{editor}{\bibfnamefont{L.~A.} \bibnamefont{Goldberg}},
  \bibinfo{editor}{\bibfnamefont{M.~M.} \bibnamefont{Halld{\'o}rsson}},
  \bibinfo{editor}{\bibfnamefont{A.}~\bibnamefont{Ing{\'o}lfsd{\'o}ttir}},
  \bibnamefont{and}
  \bibinfo{editor}{\bibfnamefont{I.}~\bibnamefont{Walukiewicz}}
  (\bibinfo{publisher}{Springer}, \bibinfo{year}{2008}), vol.
  \bibinfo{volume}{5125} of \emph{\bibinfo{series}{Lecture Notes in Computer
  Science}}, pp. \bibinfo{pages}{133--144}, ISBN
  \bibinfo{isbn}{978-3-540-70574-1}.

\bibitem[{\citenamefont{Kleinberg}(2006)}]{kleinberg06-review-pnas}
\bibinfo{author}{\bibfnamefont{J.}~\bibnamefont{Kleinberg}},
  \bibinfo{journal}{Proc. Int. Congr. Math. (ICM)}
  \textbf{\bibinfo{volume}{3}}, \bibinfo{pages}{1019} (\bibinfo{year}{2006}).

\bibitem[{\citenamefont{Mahadevan et~al.}(2006)\citenamefont{Mahadevan,
  Krioukov, Fomenkov, Huffaker, Dimitropoulos, kc~claffy, and
  Vahdat}}]{MaKrFo06}
\bibinfo{author}{\bibfnamefont{P.}~\bibnamefont{Mahadevan}},
  \bibinfo{author}{\bibfnamefont{D.}~\bibnamefont{Krioukov}},
  \bibinfo{author}{\bibfnamefont{M.}~\bibnamefont{Fomenkov}},
  \bibinfo{author}{\bibfnamefont{B.}~\bibnamefont{Huffaker}},
  \bibinfo{author}{\bibfnamefont{X.}~\bibnamefont{Dimitropoulos}},
  \bibinfo{author}{\bibnamefont{kc~claffy}}, \bibnamefont{and}
  \bibinfo{author}{\bibfnamefont{A.}~\bibnamefont{Vahdat}},
  \bibinfo{journal}{Comput Commun Rev} \textbf{\bibinfo{volume}{36}},
  \bibinfo{pages}{17} (\bibinfo{year}{2006}).

\bibitem[{\citenamefont{Bogu{\~n}\'{a}
  et~al.}(2004{\natexlab{a}})\citenamefont{Bogu{\~n}\'{a}, Pastor-Satorras,
  D\'{i}az-Guilera, and Arenas}}]{BoPa04a}
\bibinfo{author}{\bibfnamefont{M.}~\bibnamefont{Bogu{\~n}\'{a}}},
  \bibinfo{author}{\bibfnamefont{R.}~\bibnamefont{Pastor-Satorras}},
  \bibinfo{author}{\bibfnamefont{A.}~\bibnamefont{D\'{i}az-Guilera}},
  \bibnamefont{and} \bibinfo{author}{\bibfnamefont{A.}~\bibnamefont{Arenas}},
  \bibinfo{journal}{Phys Rev E} \textbf{\bibinfo{volume}{70}},
  \bibinfo{pages}{056122} (\bibinfo{year}{2004}{\natexlab{a}}).

\bibitem[{\citenamefont{Jeong et~al.}(2000)\citenamefont{Jeong, Tombor, Albert,
  Oltvai, and Barab\'{a}si}}]{JeToAlOlBa00}
\bibinfo{author}{\bibfnamefont{H.}~\bibnamefont{Jeong}},
  \bibinfo{author}{\bibfnamefont{B.}~\bibnamefont{Tombor}},
  \bibinfo{author}{\bibfnamefont{R.}~\bibnamefont{Albert}},
  \bibinfo{author}{\bibfnamefont{Z.~N.} \bibnamefont{Oltvai}},
  \bibnamefont{and} \bibinfo{author}{\bibfnamefont{A.-L.}
  \bibnamefont{Barab\'{a}si}}, \bibinfo{journal}{Nature}
  \textbf{\bibinfo{volume}{407}}, \bibinfo{pages}{651} (\bibinfo{year}{2000}).

\bibitem[{\citenamefont{Barrat et~al.}(2004)\citenamefont{Barrat,
  Barth\'{e}lemy, Pastor-Satorras, and Vespignani}}]{Barrat:2004b}
\bibinfo{author}{\bibfnamefont{A.}~\bibnamefont{Barrat}},
  \bibinfo{author}{\bibfnamefont{M.}~\bibnamefont{Barth\'{e}lemy}},
  \bibinfo{author}{\bibfnamefont{R.}~\bibnamefont{Pastor-Satorras}},
  \bibnamefont{and}
  \bibinfo{author}{\bibfnamefont{A.}~\bibnamefont{Vespignani}},
  \bibinfo{journal}{Proc. Natl. Acad. Sci. USA} \textbf{\bibinfo{volume}{101}},
  \bibinfo{pages}{3747} (\bibinfo{year}{2004}).

\bibitem[{\citenamefont{Amaral et~al.}(2000)\citenamefont{Amaral, Scala,
  Barth\'{e}lemy, and Stanley}}]{Amaral:2000}
\bibinfo{author}{\bibfnamefont{L.~A.~N.} \bibnamefont{Amaral}},
  \bibinfo{author}{\bibfnamefont{A.}~\bibnamefont{Scala}},
  \bibinfo{author}{\bibfnamefont{M.}~\bibnamefont{Barth\'{e}lemy}},
  \bibnamefont{and} \bibinfo{author}{\bibfnamefont{H.~E.}
  \bibnamefont{Stanley}}, \bibinfo{journal}{Proc. Nat. Acad. Sci.}
  \textbf{\bibinfo{volume}{97}}, \bibinfo{pages}{11149} (\bibinfo{year}{2000}).

\bibitem[{USA()}]{USAN}
\bibinfo{note}{Data available at http://www.transtats.bts.gov/}.

\bibitem[{\citenamefont{Meyer et~al.}(2007)\citenamefont{Meyer, Zhang, and
  Fall}}]{iab-raws-report-phys}
\bibinfo{editor}{\bibfnamefont{D.}~\bibnamefont{Meyer}},
  \bibinfo{editor}{\bibfnamefont{L.}~\bibnamefont{Zhang}}, \bibnamefont{and}
  \bibinfo{editor}{\bibfnamefont{K.}~\bibnamefont{Fall}}, eds.,
  \emph{\bibinfo{title}{RFC4984}} (\bibinfo{publisher}{The Internet
  Architecture Board}, \bibinfo{year}{2007}).

\bibitem[{\citenamefont{Krioukov et~al.}(2007)\citenamefont{Krioukov,
  kc~claffy, Fall, and Brady}}]{KrKc07}
\bibinfo{author}{\bibfnamefont{D.}~\bibnamefont{Krioukov}},
  \bibinfo{author}{\bibnamefont{kc~claffy}},
  \bibinfo{author}{\bibfnamefont{K.}~\bibnamefont{Fall}}, \bibnamefont{and}
  \bibinfo{author}{\bibfnamefont{A.}~\bibnamefont{Brady}},
  \bibinfo{journal}{Comput Commun Rev} \textbf{\bibinfo{volume}{37}},
  \bibinfo{pages}{41} (\bibinfo{year}{2007}).

\bibitem[{\citenamefont{Ravasz et~al.}(2007)\citenamefont{Ravasz, Gnanakaran,
  and Toroczkai}}]{RaGna07}
\bibinfo{author}{\bibfnamefont{E.}~\bibnamefont{Ravasz}},
  \bibinfo{author}{\bibfnamefont{S.}~\bibnamefont{Gnanakaran}},
  \bibnamefont{and}
  \bibinfo{author}{\bibfnamefont{Z.}~\bibnamefont{Toroczkai}},
  \emph{\bibinfo{title}{Network structure of protein folding pathways}}
  (\bibinfo{year}{2007}), \bibinfo{note}{\url{arXiv:0705.0912}}.

\bibitem[{\citenamefont{Krioukov et~al.}(2008)\citenamefont{Krioukov,
  Papadopoulos, Bogu{\~{n}}\'{a}, and Vahdat}}]{KrPa08}
\bibinfo{author}{\bibfnamefont{D.}~\bibnamefont{Krioukov}},
  \bibinfo{author}{\bibfnamefont{F.}~\bibnamefont{Papadopoulos}},
  \bibinfo{author}{\bibfnamefont{M.}~\bibnamefont{Bogu{\~{n}}\'{a}}},
  \bibnamefont{and} \bibinfo{author}{\bibfnamefont{A.}~\bibnamefont{Vahdat}},
  \emph{\bibinfo{title}{Efficient navigation in scale-free networks embedded in
  hyperbolic metric spaces}} (\bibinfo{year}{2008}),
  \bibinfo{note}{\url{arXiv:0805.1266}}.

\bibitem[{\citenamefont{Bogu{\~{n}}\'{a} and Pastor-Satorras}(2003)}]{BoPa03}
\bibinfo{author}{\bibfnamefont{M.}~\bibnamefont{Bogu{\~{n}}\'{a}}}
  \bibnamefont{and}
  \bibinfo{author}{\bibfnamefont{R.}~\bibnamefont{Pastor-Satorras}},
  \bibinfo{journal}{Phys Rev E} \textbf{\bibinfo{volume}{68}},
  \bibinfo{pages}{036112} (\bibinfo{year}{2003}).

\bibitem[{\citenamefont{{M. Giot {\it et al.}}}(2003)}]{Giot03}
\bibinfo{author}{\bibnamefont{{M. Giot {\it et al.}}}},
  \bibinfo{journal}{Science} \textbf{\bibinfo{volume}{302}},
  \bibinfo{pages}{1727} (\bibinfo{year}{2003}).

\bibitem[{\citenamefont{Shavitt and Shir}(2005)}]{dimes-ccr}
\bibinfo{author}{\bibfnamefont{Y.}~\bibnamefont{Shavitt}} \bibnamefont{and}
  \bibinfo{author}{\bibfnamefont{E.}~\bibnamefont{Shir}},
  \bibinfo{journal}{Comput Commun Rev} \textbf{\bibinfo{volume}{35}}
  (\bibinfo{year}{2005}).

\bibitem[{\citenamefont{Park and Newman}(2003)}]{Park:2003}
\bibinfo{author}{\bibfnamefont{J.}~\bibnamefont{Park}} \bibnamefont{and}
  \bibinfo{author}{\bibfnamefont{M.~E.~J.} \bibnamefont{Newman}},
  \bibinfo{journal}{Phys. Rev. E} \textbf{\bibinfo{volume}{68}},
  \bibinfo{pages}{026112} (\bibinfo{year}{2003}).

\bibitem[{\citenamefont{Bogu{\~n}\'{a}
  et~al.}(2004{\natexlab{b}})\citenamefont{Bogu{\~n}\'{a}, Pastor-Satorras, and
  Vespignani}}]{Boguna:2004}
\bibinfo{author}{\bibfnamefont{M.}~\bibnamefont{Bogu{\~n}\'{a}}},
  \bibinfo{author}{\bibfnamefont{R.}~\bibnamefont{Pastor-Satorras}},
  \bibnamefont{and}
  \bibinfo{author}{\bibfnamefont{A.}~\bibnamefont{Vespignani}},
  \bibinfo{journal}{European Physical Journal B} \textbf{\bibinfo{volume}{38}},
  \bibinfo{pages}{205} (\bibinfo{year}{2004}{\natexlab{b}}).

\bibitem[{\citenamefont{Alvarez-Hamelin
  et~al.}(2006)\citenamefont{Alvarez-Hamelin, Dall'Asta, Barrat, and
  Vespignani}}]{Lanet-vi}
\bibinfo{author}{\bibfnamefont{J.~I.} \bibnamefont{Alvarez-Hamelin}},
  \bibinfo{author}{\bibfnamefont{L.}~\bibnamefont{Dall'Asta}},
  \bibinfo{author}{\bibfnamefont{A.}~\bibnamefont{Barrat}}, \bibnamefont{and}
  \bibinfo{author}{\bibfnamefont{A.}~\bibnamefont{Vespignani}}, in
  \emph{\bibinfo{booktitle}{Advances in Neural Information Processing Systems
  18}}, edited by \bibinfo{editor}{\bibfnamefont{Y.}~\bibnamefont{Weiss}},
  \bibinfo{editor}{\bibfnamefont{B.}~\bibnamefont{Sch\"{o}lkopf}},
  \bibnamefont{and} \bibinfo{editor}{\bibfnamefont{J.}~\bibnamefont{Platt}}
  (\bibinfo{publisher}{MIT Press}, \bibinfo{address}{Cambridge, MA},
  \bibinfo{year}{2006}), pp. \bibinfo{pages}{41--50}.

\bibitem[{\citenamefont{Bollob{\'a}s}(1998)}]{bollobas98}
\bibinfo{author}{\bibfnamefont{B.}~\bibnamefont{Bollob{\'a}s}},
  \emph{\bibinfo{title}{Modern Graph Theory}}
  (\bibinfo{publisher}{Springer-Verlag}, \bibinfo{address}{New York},
  \bibinfo{year}{1998}).

\bibitem[{\citenamefont{Dorogovtsev et~al.}(2006)\citenamefont{Dorogovtsev,
  Goltsev, and Mendes}}]{DoGo06}
\bibinfo{author}{\bibfnamefont{S.~N.} \bibnamefont{Dorogovtsev}},
  \bibinfo{author}{\bibfnamefont{A.~V.} \bibnamefont{Goltsev}},
  \bibnamefont{and} \bibinfo{author}{\bibfnamefont{J.~F.~F.}
  \bibnamefont{Mendes}}, \bibinfo{journal}{Phys Rev Lett}
  \textbf{\bibinfo{volume}{96}}, \bibinfo{pages}{040601}
  (\bibinfo{year}{2006}).

\end{thebibliography}
\end{document}